\documentclass[nofootinbib,superscriptaddress,aps]{revtex4}

\usepackage{graphicx,bm,color}

\newcommand{\be}{\begin{equation}}
\newcommand{\ee}{\end{equation}}
\newcommand{\bea}{\begin{eqnarray}}
\newcommand{\eea}{\end{eqnarray}}
\newcommand{\pup}{p^\uparrow}
\newcommand{\qup}{q^\uparrow}

\begin{document}

\title{Transverse single-spin asymmetries in $\ell \,p^\uparrow\to h \,X$  within a TMD approach:\\
Role of quasireal photon exchange}

\author{Umberto D'Alesio}
\email{umberto.dalesio@ca.infn.it}
\affiliation{Dipartimento di Fisica, Universit\`a di Cagliari, Cittadella Universitaria, I-09042 Monserrato (CA), Italy}
\affiliation{INFN, Sezione di Cagliari, Cittadella Universitaria, I-09042 Monserrato (CA), Italy}

\author{Carlo Flore}
\email{carlo.flore@ca.infn.it}
\affiliation{Dipartimento di Fisica, Universit\`a di Cagliari, Cittadella Universitaria, I-09042 Monserrato (CA), Italy}
\affiliation{INFN, Sezione di Cagliari, Cittadella Universitaria, I-09042 Monserrato (CA), Italy}

\author{Francesco Murgia}
\email{francesco.murgia@ca.infn.it}
\affiliation{INFN, Sezione di Cagliari, Cittadella Universitaria, I-09042 Monserrato (CA), Italy}

\begin{abstract}
We present an updated study of transverse single-spin asymmetries for the inclusive large-$P_T$ processes $\ell \, p^\uparrow \to h\, X$ and $\ell\, p^\uparrow \to {\rm jet}\,X$, within a transverse momentum-dependent approach, including the contribution of quasireal (Weizs\"acker-Williams) photons. In the spirit of a unified transverse momentum-dependent scheme, predictions are obtained adopting the Sivers and transversity distributions and the Collins fragmentation functions as extracted from fits to the azimuthal asymmetries measured in semi-inclusive deep inelastic scattering and $e^+e^-$ annihilation processes. The description of the available data is extremely good, showing a clear general improvement with respect to the previous leading-order analysis. Predictions for unpolarized cross sections and single-spin asymmetries for  ongoing and future experiments are also given.
\end{abstract}


\maketitle

\section{Introduction}
\label{intro}

The role played by transverse single-spin asymmetries (SSAs) in our understanding of the nucleon structure is nowadays well consolidated and, at the same time, still source of challenging issues. Indeed, SSAs observed in processes where two energy scales (a large and a small one) are detected are unambiguously studied within an approach based on factorization theorems in terms of transverse momentum dependent distributions (TMDs). On the other hand, the description of the large data sets for the SSA $A_N$ measured in inclusive pion production in $p^\uparrow p$ collisions, where only one energy scale is present, is still under debate (see for instance Refs.~\cite{D'Alesio:2007jt,Aschenauer:2015ndk} for general overviews, and Refs.~\cite{Adams:1991rw,Adams:1991cs,Adams:1991rv,Adams:1991ru,Adams:2003fx,Adler:2005in,Lee:2007zzh, Abelev:2008af,Adamczyk:2012xd,Igo:2012,Bland:2013pkt} for the experimental results).

In Refs.~\cite{Anselmino:2009pn,Anselmino:2014eza} this issue was investigated in a somehow theoretically more simple  single-inclusive process, $\ell \, \pup \to h \, X$,  still characterized by a single large energy scale, but very close to the semi-inclusive deep inelastic scattering (SIDIS) process, for which TMD factorization has been proved~\cite{Collins:2002kn, Collins:2004nx,Ji:2004wu, Ji:2004xq, Bacchetta:2008xw,Collins:2011zzd,GarciaEchevarria:2011rb,Echevarria:2014rua}.

This process indeed can be considered a sort of bridge between the $p^\uparrow p\to h\,X$ and $\ell\,p^\uparrow\to \ell' h\, X$ processes: it is single inclusive with a single large energy scale (as the $pp\to h\,X$ process), and at the same time, at leading order, is controlled by the colour-blind electromagnetic interaction (as the SIDIS process). This should reduce the role played by initial/final state interactions leading to potential factorization breaking effects. On the other hand, adopting the relevant TMDs (Sivers and Collins functions), as extracted from SIDIS data, in the inclusive hadron production in lepton-proton collisions represents an attempt towards (and a test for) a unified TMD scheme. It is worth mentioning that the same process was also considered in Refs.~\cite{Koike:2002gm,Gamberg:2014eia} in the framework of collinear factorization with twist-three correlation functions, while inclusive jet production was studied in Ref.~\cite{Kang:2011jw}.

In Refs.~\cite{Anselmino:2009pn,Anselmino:2014eza}, to which we refer the reader for all details of the approach,
SSAs were computed assuming a TMD factorization scheme at leading order (LO), that is considering only the elementary partonic channel $\ell q\to \ell q$.
In particular, in Ref.~\cite{Anselmino:2014eza} the theoretical estimates were compared with a selection\footnote{Only data for inclusive events in the backward target hemisphere at large $P_T$ and tagged events (deep inelastic scattering category) were considered.} of the experimental results by the HERMES Collaboration~\cite{Airapetian:2013bim}, showing a good agreement in sign and size. In spite of this, it was also pointed out that some of the discrepancies still present between theory and experiment could be ascribed to effects neglected in a LO treatment.

Here we want to extend this LO study including the contribution from quasireal photon exchange, in the Weizs\"acker-Williams approximation, potentially relevant in the kinematical configuration dominated by small $Q^2$. This will allow us, still within a TMD scheme, to improve the description of the fully-inclusive data and consider, for the first time, the HERMES anti-tagged data set, dominated by events in which the final lepton (not observed) has a very small scattering angle. Notice that this data category was not included in the previous analysis because a simple LO approach (namely via $\ell q\to \ell q$) is expected to be not adequate.

In this respect we will benefit from the study performed in Ref.~\cite{Hinderer:2015hra}, even if with a different perspective and approach. In this work the authors, within a collinear-factorization scheme, computed the next-to-leading order (NLO) corrections to the unpolarized cross sections for the same process and discussed the role of quasireal photon exchange. In most kinematical configurations they found that this contribution represents only a small part of the NLO corrections. They then concluded that only a full NLO treatment could be considered complete.

On the other hand, within a TMD scheme, as well as in the twist-three approach, NLO corrections are still not available for such a process and it is then worth seeing to what extent the quasireal photon exchange could play a role in the computation of spin asymmetries. On top of that, and relevant from our point of view, by including transverse momentum effects the estimates of unpolarized cross sections are enhanced w.r.t.~those computed in a collinear framework. Experimental data, still not available, would definitely help in this respect. Notice that for the process $pp\to\pi \,X$ the estimates of unpolarized cross sections in a TMD approach at leading order show a reasonable agreement with available data from the Relativistic Heavy Ion Collider (RHIC); see Ref.~\cite{Boglione:2007dm}.

The main aim of this study will be then to provide the complete calculation within a TMD formalism of the quasireal photon exchange in $\ell\, p\to h\, X$ and $\ell\, p\to {\rm jet} \,X$ processes and to compute the unpolarized cross sections and the SSAs for various experimental setups.

The paper is organized as follows: in Section~\ref{form} we recall the general formalism, deriving and discussing all new theoretical results. In particular, in Section~\ref{formWW} we present, for the first time, the full TMD expressions for the quasireal photon contribution to unpolarized and transversely polarized cross sections for inclusive hadron and inclusive jet production. In Section~\ref{res} we show our phenomenological results, starting with the unpolarized cross sections for HERMES, Jefferson Lab (JLab), COMPASS and Electron-Ion Collider (EIC) experiments, and then focusing on transverse SSAs, with special emphasis on the comparison with HERMES data. Predictions for other experimental setups are also given and discussed. Conclusions and final comments are gathered in Section~\ref{concl}.

\section{Formalism}
\label{form}

We consider the transverse single-spin asymmetry, $A_N$, for the process $\pup \ell \to h \, X$ in the proton-lepton center-of-mass ({\it c.m.}) frame,
 \be
A_N = \frac{d\sigma^\uparrow(\bm{P}_T) - d\sigma^\downarrow(\bm{P}_T)}
           {d\sigma^\uparrow(\bm{P}_T) + d\sigma^\downarrow(\bm{P}_T)}
    = \frac{d\sigma^\uparrow(\bm{P}_T) - d\sigma^\uparrow(-\bm{P}_T)}
           {2 \, d\sigma^{\rm unp}(\bm{P}_T)} = \frac{d\Delta \sigma(\bm{P}_T)}{2 \, d\sigma^{\rm unp}(\bm{P}_T)}\,, \label{an}
\ee
where
\be
d\sigma^{\uparrow, \downarrow} \equiv \frac{E_h \, d\sigma^{p^{\uparrow,
\downarrow} \, \ell \to h\, X}}{d^{3} \bm{P}_h}
\ee
and $\bm{P}_h$ and $\bm{P}_T$ are respectively the three-momentum of the final hadron and its vector transverse component. The polarized proton (or nucleon) is in a pure transverse spin state $\bm{S}$ and is assumed to move along the positive $Z_{\rm cm}$ axis, while the lepton is taken unpolarized. We define as transverse polarization for the proton the $Y_{\rm cm}$ direction, with $\uparrow$ and $\downarrow$ respectively for protons polarized along or opposite to $Y_{\rm cm}$. The $X_{\rm cm}$ axis is defined in such a way that a hadron $h$ with $(P_h)_{X_{\rm cm}} > 0$ is produced {\it to the left} of the incoming proton (see also Fig.~1 of Ref.~\cite{Anselmino:2009pn}).

Notice that for a generic transverse polarization, $\bm{S}_T$, along an azimuthal direction $\phi_S$ in the chosen reference frame, in which the $\uparrow$ direction is given by $\phi_S = \pi/2$, one has:
\be
A(\phi_S, S_T) = \bm{S}_T \cdot (\hat{\bm{p}} \times \hat{\bm{P}}_T) \, A_N = S_T \sin\phi_S \, A_N \>, \label{phis}
\ee
where $\bm{p}$ is the proton momentum.
Following the usual definition adopted in SIDIS experiments, one simply obtains:
\be
A_{TU}^{\sin\phi_S} \equiv \frac{2}{S_T} \,
\frac{\int \, d\phi_S \> [d\sigma(\phi_S) - d\sigma(\phi_S + \pi)]\> \sin\phi_S}
     {\int \, d\phi_S \> [d\sigma(\phi_S) + d\sigma(\phi_S + \pi)]}
= A_N \>.
\label{ATU}
\ee

In order to include effects from quasireal photon exchange, adopting the Weizs\"acker-Williams (WW) approximation, within a TMD approach, we write the SSA under consideration as follows:
\be
\label{ANWW}
A_N = \frac{d\Delta\sigma^{\rm LO} + d\Delta\sigma^{\rm WW}}{2[ d\sigma^{\rm LO} + d\sigma^{\rm WW}]}\,,
\ee
where the leading-order contributions are given by~\cite{Anselmino:2009pn,Anselmino:2014eza},
\bea
d\Delta\sigma^{\rm LO} &=& {\displaystyle \sum_{q} \int \frac{dx \, dz}
{16\,\pi^2 x\,z^2 s}}\;
d^2 \bm{k}_{\perp} \, d^3 \bm{p}_{\perp}\,
\delta(\bm{p}_{\perp} \cdot \hat{\bm{p}}'_q) \, J(p_\perp)
\> \delta(\hat s + \hat t + \hat u)
\> [\Sigma(\uparrow) - \Sigma(\downarrow)]^{q \ell \to q \ell}\nonumber\\
&&\\
2\,d\sigma^{\rm LO} &=& {\displaystyle \sum_{q} \int \frac{dx \, dz}
{16\,\pi^2 x\,z^2 s}}\;
d^2 \bm{k}_{\perp} \, d^3 \bm{p}_{\perp}\,
\delta(\bm{p}_{\perp} \cdot \hat{\bm{p}}'_q) \, J(p_\perp)
\> \delta(\hat s + \hat t + \hat u)
\> [\Sigma(\uparrow) + \Sigma(\downarrow)]^{q \ell \to q \ell}\nonumber\,,\\
&&
\eea
with $q=u,\bar u, d, \bar d, s, \bar s$ and
\bea
\,[\Sigma(\uparrow) - \Sigma(\downarrow)]^{q \ell \to q \ell} &=&
\frac{1}{2} \, \Delta^N\! f_{q/\pup}(x,k_{\perp}) \cos\phi \, \left[\,|{\hat M}_1^0|^2 + |{\hat M}_2^0|^2 \right] \, D_{h/q} (z, p_{\perp})  \nonumber \\
&+& h_{1q}(x,k_{\perp}) \, \hat M_1^0 \hat M_2^0 \, \Delta^N\! D_{h/\qup} (z, p_{\perp}) \, \cos(\phi' + \phi_q^h) \nonumber\\
&-& \frac{k_\perp^2}{2M^2} h_{1T}^{\perp q}(x,k_{\perp}) \,\hat M_1^0 \hat M_2^0 \, \Delta^N\!D_{h/\qup}(z, p_{\perp}) \, \cos(2\phi - \phi' - \phi_q^h)
\label{ds1}\\
\,[\Sigma(\uparrow) + \Sigma(\downarrow)]^{q \ell \to q \ell} & = &
f_{q/p} (x,k_{\perp}) \, \left[\,|{\hat M}_1^0|^2 + |{\hat M}_2^0|^2 \right] \, D_{h/q} (z, p_{\perp}) \nonumber\\
\label{ss1}
&-& \frac{k_\perp}{M} \, h_{1}^{\perp q}(x,k_{\perp}) \, \hat M_1^0 \hat M_2^0 \, \Delta^N\!D_{h/\qup}(z, p_{\perp}) \, \cos(\phi- \phi' - \phi_q^h)\,.
\eea
Proper definition of all functions and variables appearing in the above equations can be found in Ref.~\cite{Anselmino:2009pn} and its Appendices and in Ref.~\cite{Anselmino:2005sh}. For a better understanding we recall here their physical meaning.
\begin{itemize}
\item
$\bm{k}_\perp = k_\perp\,(\cos\phi, \sin\phi, 0)$ and $\bm{p}_\perp$ are respectively the transverse momentum of the parton in the proton and of the final hadron with respect to the direction of the fragmenting parent parton, with momentum $\bm{p}^\prime_q$. Notice that $\bm{p}$ and $\bm{p}_\perp$ are different vectors.
\item
The first term on the r.h.s.~of Eq.~(\ref{ds1}) represents the Sivers effect~\cite{Sivers:1989cc,Sivers:1990fh,Bacchetta:2004jz}, with
\bea
\Delta \hat f_{q/p,S}(x, \bm{k}_{\perp}) &=& \hat f_{q/p,S}(x,\bm{k}_{\perp}) - \hat f_{q/p,-S}(x, \bm{k}_{\perp})\nonumber\\
&\equiv& \Delta^N\! f_{q/p^\uparrow}\,(x, k_{\perp}) \>
\hat{\bm{S}}_T \cdot (\hat{\bm{p}} \times \hat{\bm{k}}_{\perp }) \nonumber  \\
&=& -2 \, \frac{k_\perp}{M} \, f_{1T}^{\perp q}(x, k_{\perp}) \>
\hat{\bm{S}}_T \cdot (\hat{\bm{p}} \times \hat{\bm{k}}_{\perp }) \>. \label{defsivnoi}
\eea
The extra factors are the unpolarized elementary interaction ($\propto (|M_1^0|^2 + |M_2^0|^2)$) and the unpolarized fragmentation function $D_{h/q} (z, p_{\perp})$; in the chosen reference frame, where $\phi_S=\pi/2$, the correlation factor $\hat{\bm{S}}_T \cdot (\hat{\bm{p}} \times
\hat{\bm{k}}_{\perp })$ gives the modulation $\sin(\phi_S-\phi)=\cos\phi$.
\item
The second and third terms (this last one numerically negligible) on the r.h.s.~of Eq.~(\ref{ds1}) represent the contribution to $A_N$ of the Collins effect, given respectively as a convolution of the unintegrated transversity distribution, $h_{1q}(x,k_{\perp})$, and the pretzelosity distribution, $h_{1T}^{\perp q}(x,k_{\perp})$, with the Collins function $\Delta^N\! D_{h/q^\uparrow} (z, p_{\perp})$~\cite{Collins:1992kk,Bacchetta:2004jz},
\bea
\Delta \hat D_{h/q^\uparrow}\,(z, \bm{p}_{\perp}) &=& \hat D_{h/q^\uparrow}\,(z, \bm{p}_{\perp}) - \hat D_{h/q^\downarrow}\,(z,\bm{p}_{\perp}) \nonumber\\
&\equiv& \Delta^N\! D_{h/q^\uparrow}\,(z, p_{\perp}) \>
\hat{\bm{s}}_q \cdot (\hat{\bm{p}}_q^\prime \times \hat{\bm{p}}_{\perp }) \nonumber \\
\label{defcolnoi} &=& \frac{2 \, p_\perp}{z \, m_h} H_{1}^{\perp q}(z,
p_{\perp}) \> \hat{\bm{s}}_q \cdot (\hat{\bm{p}}_q^\prime \times
\hat{\bm{p}}_{\perp}) \>.
\eea
The product $\hat M_1^0 \, \hat M_2^0$ is related to the spin transfer elementary interaction ($\propto d\hat\sigma^{q^\uparrow \ell \to q^\uparrow \ell} - d\hat\sigma^{q^\uparrow \ell \to q^\downarrow \ell})$, while the factors $\cos(\phi' + \phi_q^h)$ and $\cos(2\phi - \phi' - \phi_q^h)$ arise from phases in the $\bm{k}_\perp$-dependent transversity and pretzelosity distributions, the Collins function and the elementary polarized interaction.
\item
The first (and dominant) term on the r.h.s.~of  Eq.~(\ref{ss1}) is the convolution of the unpolarized TMD parton distribution and fragmentation functions with the unpolarized partonic interactions, while the second one, numerically negligible, represents the Boer-Mulders mechanism~\cite{Boer:1997nt,Boer:1999mm}, with the corresponding function defined as
\bea
\Delta \hat f_{q,s/p}(x, \bm{k}_{\perp}) &=& \hat f_{q,s/p}(x,\bm{k}_{\perp}) - \hat f_{q,-s/p}(x, \bm{k}_{\perp}) \nonumber\\
&\equiv& \Delta^N\! f_{q^\uparrow/p}\,(x, k_{\perp}) \> \hat{\bm{s}}_T \cdot (\hat{\bm{p}} \times \hat{\bm{k}}_{\perp }) \nonumber       \\
\label{defboernoi}
&=& -\, \frac{k_\perp}{M} \, h_{1}^{\perp q}(x, k_{\perp}) \>
\hat{\bm{s}}_T \cdot (\hat{\bm{p}} \times \hat{\bm{k}}_{\perp }) \>.
\eea
\end{itemize}

In the following Sections we discuss in detail the Weizs\"acker-Williams approximation and its role in the (un)polarized process under consideration.

\subsection{Weizs\"acker-Williams approximation}

As shown in Ref.~\cite{Hinderer:2015hra}, in a NLO treatment of the inclusive process $\ell\, p\to h\,X$, the collinear lepton singularities could be regularized, and opportunely redefined, by introducing a QED parton distribution for the lepton, in strong analogy with the ordinary nucleon's parton distributions. The only difference is that in such a case the partons are the lepton itself and the photon. Without entering into many details, we can say that at order $\alpha^2\alpha_s$ there will be a contribution from the photon acting as a parton of the lepton and entering the hard scattering process. This can be represented as a Weizs\"acker-Williams contribution~\cite{vonWeizsacker:1934nji,Williams:1934ad}, where the lepton acts as a source of real photons (see also Refs.~\cite{Brodsky:1971ud,Terazawa:1973tb,Kniehl:1990iv}). We then assume the following factorization formula for the WW contribution to the process $\ell\, p \to h\, X$:
\be
\label{WW}
\sigma^{\rm WW}(\ell p \to h\, X) = \int d y f_{\gamma/\ell}(y)\, \sigma(\gamma p\to h\, X) \,,
\ee
where $f_{\gamma/\ell}(y)$ is the number density of photons inside the lepton, carrying a lepton-momentum fraction $y$ ($p_\gamma = y p_\ell$) and $\sigma(\gamma p\to h \,X)$ is the cross section for the process $\gamma p\to h \,X$ initiated by a real photon.

For the WW distribution we follow Ref.~\cite{Hinderer:2015hra}, adopting
\be
f_{\gamma/\ell}(y) = \frac{\alpha}{2\pi} \frac{1+(1-y)^2}{y}\Biggl[\ln \Biggl({\frac{\mu^2}{y^2 m^2_{\ell}}}\Biggr) - 1 \Biggr] + \mathcal{O}(\alpha^2)
  \,,
\label{ww}
\ee
where $\alpha$ is the electromagnetic coupling constant, $\mu$ the factorization scale and $m_\ell$ the lepton mass. We have also tried an alternative form for the WW distribution, like the one proposed in Refs.~\cite{Brodsky:1971ud,Terazawa:1973tb,Kniehl:1990iv} and adopted, in the context of SSA studies, in Refs.~\cite{Godbole:2012bx,Godbole:2013bca}. In both cases we have considered two choices of the factorization scale, namely $\mu=P_T$ or $\mu=\sqrt s/2$. Since these choices do not lead to any significant differences we will present our estimates only for the form in Eq.~(\ref{ww}) with $\mu=P_T$.

\subsection{Quasireal photon contribution to SSAs for inclusive particle production}
\label{formWW}

In order to compute the WW contribution to $A_N$, based on the factorized expression~(\ref{WW}), we start with the general treatment for the cross section, in a TMD scheme, of the large-$P_T$ inclusive polarized process $A(S_A) \, B(S_B) \to C \, X$~\cite{Anselmino:2005sh}, adapted here to the process $p(S)\,\ell\to h\, X$:
\bea
\label{master}
    \frac{E_hd\sigma^{p(S)\,\ell\to h\, X}_{\rm WW}}{d^3\bm{P}_h}
	    & = & \sum_{a,c,d,\{\lambda\}}\int\!\frac{dx\, dy\, dz}{16\pi^2 x y z^2 s} d^2\!\bm{k}_\perp\,
d^3\!\bm{p}_{\perp}\, \delta(\bm{p}_{\perp}\cdot \hat{\bm{p}_c})\, J(p_\perp)\,\delta(\hat s +\hat t+ \hat u)  \nonumber\\
& \times &  \rho_{\lambda _a\lambda_{a}'}^{a/p,S}\, \hat f_{a/p,S}(x,\bm{k}_{\perp})\,\rho_{\lambda _\gamma\lambda_{\gamma}'}^{\gamma/\ell}\, f_{\gamma/\ell}(y)
\hat M_{\lambda_c,\lambda_d;\lambda_a\lambda_\gamma}\,\hat M_{\lambda_c',\lambda_d;\lambda_a'\lambda_\gamma'}^* \, D_{\lambda_c,\lambda_c'}^{\lambda_h,\lambda_h}(z,\bm{p}_\perp)\,,\nonumber\\
&&
\eea
which can be written schematically as
\bea
\label{master2}
\!\!\!\!\!\!\!d\sigma^{\rm WW}(S) &=&  \sum_{a,c,d}\int\!\frac{dx\, dy\, dz}{16\pi^2 x y z^2 s} d^2\!\bm{k}_\perp
d^3\!\bm{p}_{\perp} \delta(\bm{p}_{\perp}\cdot \hat{\bm{p}_c})\, J(p_\perp)\,\delta(\hat s +\hat t+ \hat u)
\, \Sigma(S)^{a\gamma\to cd}
\,.
\eea

Notice that in Eq.~(\ref{master}) we have consistently adopted a collinear WW distribution, as properly defined for the case of a scattered lepton, and a photon, almost collinear with the initial lepton and that now $a,c$ can be a quark (antiquark) or a gluon (this is at variance w.r.t.~the LO calculation where only quark TMDs are involved).

For the notation and the meaning of the quantities entering Eq.~(\ref{master}) we refer the reader to Refs.~\cite{Anselmino:2009pn,Anselmino:2005sh}. Here we only note that the Mandelstam variables for the process $a\gamma\to c d$ are defined using $p_\gamma = y p_\ell$ and that the $\rho$'s and the $\hat M$'s are respectively the helicity density matrices of partons (photons) inside a polarized hadron (an unpolarized lepton) and the helicity amplitudes for the elementary processes $q\gamma\to q g$ and $g\gamma\to q\bar q$. We further recall that the $\hat M$'s are defined in the proton-lepton  \emph{c.m.}~frame, where the $a\gamma\to cd$ processes are not planar. They can be expressed in terms of the corresponding canonical helicity amplitudes $\hat M^0$ in the $a$-$\gamma$ \emph{c.m.}~frame by performing proper boost and rotations as described in Ref.~\cite{Anselmino:2004ky,Anselmino:2005sh} (see also Appendix~\ref{appendix}).

By summing over the helicities, using the proper definition of the helicity density matrices for spin-1/2 and spin-1 partons, and exploiting the parity properties of the helicity amplitudes, we obtain the following expressions for the kernels $\Sigma(S)^{a\gamma\to cd} $:
\begin{enumerate}
\item $q\gamma\to q g$ processes
 \bea
\Sigma(S) &=& \frac{1}{2}\,\hat f_{q/p,S}(x,\bm{k}_{\perp})\, f_{\gamma/\ell}(y) \Big\{D_{h/q}(z,p_\perp) [(|\hat{M}^0_1|^2 + |\hat{M}^0_2|^2)+
P_z^q\,P_z^\gamma \,(|\hat{M}^0_1|^2 - |\hat{M}^0_2|^2)]\nonumber \\
&-&\Delta^N\!D_{h/q^\uparrow}(z,p_\perp)\,\hat{M}^0_1 \hat{M}^0_2
[P_x^q\,\sin(\varphi_1-\varphi_2 +\phi_q^h) - P_y^q\,\cos(\varphi_1-\varphi_2 +\phi_q^h)]\Big\}\,,\label{qgaqg}
\eea
where $q$ can be either a quark or an antiquark and
\be
\label{Mqgaqg}
|\hat{M}^0_1|^2 = -\frac{16}{3}\,g_s^2 e^2 e_q^2 \,\frac{\hat s}{\hat u} \>\>\>\>\> |\hat{M}^0_2|^2 = -\frac{16}{3}\, g_s^2 e^2 e_q^2\, \frac{\hat u}{\hat s} \>\>\>\>\>  \hat{M}^0_1 \hat{M}^0_2 = \frac{16}{3}\, g_s^2  e^2 e_q^2
\ee
\item $q\gamma\to g q$ processes
\bea
\Sigma(S) &=& \frac{1}{2}\,\hat f_{q/p,S}(x,\bm{k}_{\perp})\, f_{\gamma/\ell}(y)\,\Big \{D_{h/g}(z,p_\perp)[ (|\hat{M}^0_1|^2 + |\hat{M}^0_3|^2)+
P_z^q \, P_z^\gamma \,(|\hat{M}^0_1|^2 - |\hat{M}^0_3|^2)]\nonumber \\
&+& \Delta^N\!D_{h/\mathcal{T}_1^g}(z,p_\perp)\,\hat{M}^0_1 \hat{M}^0_3
\,[\mathcal{T}_1^\gamma\,\cos(\varphi_1-\varphi_3 +2\phi_g^h) + \mathcal{T}_2^\gamma\,\sin(\varphi_1-\varphi_3 +2\phi_g^h)]\Big\}\,,\nonumber\\
&&\label{qgagq}
\eea
where again $q$ can be either a quark or an antiquark and
\be
\label{Mqgagq}
|\hat{M}^0_1|^2 = -\frac{16}{3}\,g_s^2e^2e_q^2 \,\frac{\hat s}{\hat t} \>\>\>\>\> |\hat{M}^0_3|^2 = -\frac{16}{3}\,g_s^2e^2e_q^2\, \frac{\hat t}{\hat s} \>\>\>\>\>  \hat{M}^0_1 \hat{M}^0_3 = \frac{16}{3}\,g_s^2e^2e_q^2
\ee
\item $g\gamma\to q\bar q$ processes
\bea
\Sigma(S) &=& \frac{1}{2}\, \hat f_{g/p,S}(x,\bm{k}_{\perp}) \, f_{\gamma/\ell}(y) D_{h/q}(z,p_\perp) \Big\{[(1-P_z^g\,P_z^\gamma) \, (|\hat{M}^0_2|^2 + |\hat{M}^0_3|^2)]\nonumber \\
&+& 2\,\hat{M}^0_2 \hat{M}^0_3
\,[(\mathcal{T}_1^g\,\mathcal{T}_1^\gamma + \mathcal{T}_2^g\,\mathcal{T}_2^\gamma)\,\cos(\varphi_2-\varphi_3) + (\mathcal{T}_1^g\,\mathcal{T}_2^\gamma - \mathcal{T}_2^g\,\mathcal{T}_1^\gamma)\,\sin(\varphi_2-\varphi_3)]\Big\}\,,\nonumber\\
&& \label{ggaqqb}
\eea
where
\be
\label{Mggaqqb}
|\hat{M}^0_2|^2 = 2\,g_s^2e^2e_q^2\, \frac{\hat u}{\hat t} \>\>\>\>\> |\hat{M}^0_3|^2 = 2\,g_s^2e^2e_q^2 \,\frac{\hat t}{\hat u} \>\>\>\>\>  \hat{M}^0_2 \hat{M}^0_3 = 2\,g_s^2e^2e_q^2
\ee
\item $g\gamma\to\bar q q$ processes

These can be obtained from the $g\gamma\to q\bar q$ processes by interchanging in the two above equations $\hat t$ with $\hat u$  (that is $\hat{M}^0_2 \leftrightarrow\hat{M}^0_3$ and $\varphi_2 \leftrightarrow \varphi_3$) and $D_{h/q}$ with $D_{h/\bar q}$.
\end{enumerate}

In the above equations $P_i^{q,g,\gamma}$ stand for the quark, gluon and photon polarization vector components and $\mathcal{T}_i^{g,\gamma}$ for the gluon and photon linear polarization ones, while $\varphi_i$ are the azimuthal phases of the helicity amplitudes (see Appendix~\ref{appendix} for details).

We are now ready to compute the WW contributions to $A_N$. By choosing $\phi_S=\pi/2$ in the adopted reference frame, we have
\bea
\label{sigmaWW}
d\Delta\sigma^{\rm WW} &=& {\displaystyle \sum_{a,c,d} \int \frac{dx \,dy\, dz}
{16\,\pi^2 x\,y z^2 s}}\;
d^2 \bm{k}_{\perp} \, d^3 \bm{p}_{\perp}\,
\delta(\bm{p}_{\perp} \cdot \hat{\bm{p}}'_q) \, J(p_\perp)
\> \delta(\hat s + \hat t + \hat u)
\> [\Sigma(\uparrow) - \Sigma(\downarrow)]^{a\gamma\to cd}\nonumber\\
&&\\
2\,d\sigma^{\rm WW} &=& {\displaystyle \sum_{a,c,d} \int \frac{dx \,dy\, dz}
{16\,\pi^2 x\,y z^2 s}}\;
d^2 \bm{k}_{\perp} \, d^3 \bm{p}_{\perp}\,
\delta(\bm{p}_{\perp} \cdot \hat{\bm{p}}'_q) \, J(p_\perp)
\> \delta(\hat s + \hat t + \hat u)
\> [\Sigma(\uparrow) + \Sigma(\downarrow)]^{a\gamma\to cd}\nonumber\,, \\
&&
\eea
where
\bea
\sum_{a,c,d} [\Sigma(\uparrow) \pm \Sigma(\downarrow)]^{a\gamma\to cd}
&=& [\Sigma(\uparrow) \pm \Sigma(\downarrow)]^{q \gamma\to q g} + [\Sigma(\uparrow) \pm \Sigma(\downarrow)]^{q\gamma \to g q} \nonumber\\
&+& [\Sigma(\uparrow) \pm \Sigma(\downarrow)]^{\bar q \gamma\to \bar q g} + [\Sigma(\uparrow) \pm \Sigma(\downarrow)]^{\bar q\gamma \to g \bar q} \nonumber\\
&+& [\Sigma(\uparrow) \pm \Sigma(\downarrow)]^{g \gamma \to q \bar q} + [\Sigma(\uparrow) \pm \Sigma(\downarrow)]^{g \gamma \to \bar q q}\,,\label{kernWW}
\eea
with
\bea
\,[\Sigma(\uparrow) - \Sigma(\downarrow)]^{q \gamma \to q g}
& = & f_{\gamma/\ell}(y)
\Big\{\frac{1}{2} \, \Delta^N\! f_{q/\pup}(x,k_{\perp}) \cos\phi \, \left[\,|{\hat M}_1^0|^2 + |{\hat M}_2^0|^2 \right]^{q\gamma\to q g} \,D_{h/q} (z, p_{\perp}) \nonumber \\
&+&  h_{1q}(x,k_{\perp}) \, [\hat M_1^0 \hat M_2^0]^{q\gamma\to q g} \, \Delta^N\!D_{h/\qup} (z, p_{\perp}) \, \cos(\phi' + \phi_q^h) \nonumber\\
&-&  \frac{k_\perp^2}{2M^2} h_{1T}^{\perp q}(x,k_{\perp}) \, [\hat M_1^0 \hat M_2^0]^{q\gamma\to q g} \, \Delta^N\!D_{h/\qup}(z, p_{\perp}) \, \cos(2\phi - \phi' - \phi_q^h) \Big\}\nonumber\\
&&\label{updownqgqg}\\
\,[\Sigma(\uparrow) +\Sigma(\downarrow)]^{q \gamma \to q g} & = &
f_{\gamma/\ell}(y) f_{q/p} (x,k_{\perp}) \,\left[\,|{\hat M}_1^0|^2 + |{\hat M}_2^0|^2 \right]^{q\gamma\to q g} \, D_{h/q} (z, p_{\perp}) \nonumber\\
&-&  \frac{k_\perp}{M} \,h_{1}^{\perp q}(x,k_{\perp}) \, [\hat M_1^0 \hat M_2^0]^{q\gamma\to q g} \, \Delta^N\!D_{h/\qup}(z, p_{\perp}) \, \cos(\phi -\phi' - \phi_q^h)\nonumber\\
&&\label{updownqgqg2}\\
\,[\Sigma(\uparrow) - \Sigma(\downarrow)]^{q \gamma \to g q } & = & \frac{1}{2}\, f_{\gamma/\ell}(y)
 \, \Delta^N\! f_{q/\pup}(x,k_{\perp}) \cos\phi \, \left[\,|{\hat M}_1^0|^2 + |{\hat M}_3^0|^2 \right]^{q\gamma\to  gq} \,D_{h/g} (z, p_{\perp}) \nonumber\\
&&\label{updownqggq}\\
%
\,[\Sigma(\uparrow) +\Sigma(\downarrow)]^{q \gamma \to g q} & = & f_{\gamma/\ell}(y)
f_{q/p} (x,k_{\perp}) \,\left[\,|{\hat M}_1^0|^2 + |{\hat M}_3^0|^2 \right]^{q\gamma\to gq} \,
D_{h/g} (z, p_{\perp}) \label{updownqggq2}\\
%
\,[\Sigma(\uparrow) - \Sigma(\downarrow)]^{g \gamma \to q \bar q} & = & \frac{1}{2}\,f_{\gamma/\ell}(y)
\, \Delta^N\! f_{g/\pup}(x,k_{\perp}) \cos\phi \, \left[\,|{\hat M}_2^0|^2 + |{\hat M}_3^0|^2 \right]^{g\gamma\to q\bar q} \,D_{h/q} (z, p_{\perp})\nonumber\\
&&\label{updownggqqb}\\
%
\,[\Sigma(\uparrow) +\Sigma(\downarrow)]^{g \gamma \to q\bar q} & = & f_{\gamma/\ell}(y)
f_{g/p} (x,k_{\perp}) \,\left[\,|{\hat M}_2^0|^2 + |{\hat M}_3^0|^2 \right]^{g\gamma\to q\bar q} \,
D_{h/q} (z, p_{\perp}) \>,
\label{updownggqqb2}
\eea
and once again in Eqs.~(\ref{updownqgqg})-(\ref{updownqggq2}) $q$ can be either a quark or an antiquark, while for the $g\gamma\to \bar q q$ channel one can use the last two relations replacing $D_{h/q}$ with $D_{h/\bar q}$.
In Eqs.~(\ref{updownqgqg}) and (\ref{updownqgqg2}) we have redefined $\varphi_1-\varphi_2= \phi'-\phi$, consistently, and in agreement, with the notation adopted in the LO expressions\footnote{Notice that the explicit calculation of the azimuthal phases given in Ref.~\cite{Anselmino:2009pn} leads to the same results obtained following the boost-rotation procedure described in Refs.~\cite{Anselmino:2004ky,Anselmino:2005sh}.}.

{}In Eqs.~(\ref{updownqgqg}) and (\ref{updownqggq}) we recognize the Sivers and Collins effects. Once again, as for the LO piece, the terms involving the pretzelosity in Eq.~(\ref{updownqgqg}) and the Boer-Mulders function in Eq.~(\ref{updownqgqg2}) are numerically negligible (even saturating their positivity bounds). On the other hand, at variance with the leading-order analysis, we have also a potential contribution from the gluon Sivers function (see Eq.~(\ref{updownggqqb})). Notice that all contributions from linearly polarized gluons ($\mathcal{T}^g$) appearing in 
Eq.~(\ref{ggaqqb}) disappear since they are coupled to linearly polarized photon ($\mathcal{T}^\gamma$) distributions that are identically zero for an unpolarized initial lepton.

\subsubsection{SSAs in single-inclusive jet production at large transverse momentum}
\label{jet}

Inclusive jet production in lepton-proton collisions, although more difficult to measure, could be an invaluable tool to access the Sivers effect, as the lack of any fragmentation process forbids other contributions. In Ref.~\cite{Anselmino:2009pn} this case was discussed and some results for a high-energy electron-nucleon collider were presented. In the same spirit here we extend this analysis including the quasireal photon contribution. The expressions can be directly obtained from the case of inclusive hadron production by replacing the fragmentation functions with proper Dirac delta functions. We report here the main results for the WW contribution, referring to Ref.~\cite{Anselmino:2009pn} for the LO piece. For the master formula we have
\bea
\label{master-jet}
    \frac{E_{\rm j} d\sigma^{(p,S)\,\ell\to {\rm jet}\, X}_{\rm WW}}{d^3\bm{P}_{\rm j}}
	    & = & \sum_{a,c,d,\{\lambda\}}\int\!\frac{dx\, dy}{16\pi^2 x y s} \,d^2\!\bm{k}_\perp\,\delta(\hat s +\hat t+ \hat u)  \nonumber\\
& \times &  \rho_{\lambda _a\lambda_{a}'}^{a/p,S}\, \hat f_{a/p,S}(x,\bm{k}_{\perp})\,\rho_{\lambda _\gamma\lambda_{\gamma}'}^{\gamma/\ell}\, f_{\gamma/\ell}(y)
\hat M_{\lambda_c,\lambda_d;\lambda_a\lambda_\gamma}\,\hat M_{\lambda_c,\lambda_d;\lambda_a'\lambda_\gamma'}^*\,,
\eea
while for the contributions to $A_N(\rm jet)$
\bea
\label{sigmaWW-jet}
d\Delta\sigma^{\rm WW}_{\rm jet} &=& {\displaystyle \sum_{a,c,d} \int \frac{dx \,dy}{16\,\pi^2 x\,y s}}\,
d^2 \bm{k}_{\perp} \, \delta(\hat s + \hat t + \hat u)
\> [\Sigma(\uparrow) - \Sigma(\downarrow)]^{a\gamma\to cd}_{\rm jet}\\
2\,d\sigma^{\rm WW}_{\rm jet} &=& {\displaystyle \sum_{a,c,d} \int \frac{dx \,dy}{16\,\pi^2 x\,y s}}\,
d^2 \bm{k}_{\perp} \,
 \delta(\hat s + \hat t + \hat u)
\> [\Sigma(\uparrow) + \Sigma(\downarrow)]^{a\gamma\to cd}_{\rm jet}\,,
\eea
with Eq.~(\ref{kernWW}) still valid also for jet production. For the sums and differences of the kernels we can use the same expressions as given in Eqs.(\ref{updownqgqg})-(\ref{updownggqqb2}) replacing
$D_{h/q,g}(z,p_\perp)$ with 1 and $\Delta^N\!D_{h/q^\uparrow}(z,p_\perp)$ with 0. In this case, obviously, there is no fragmentation process and only the Sivers effect contributes to $A_N$. Notice that in the present treatment the jet coincides with a single final parton.

\section{Phenomenological results, comparison with data and predictions}
\label{res}

In this Section we present our theoretical estimates of the unpolarized cross sections and the SSAs for inclusive pion production in lepton-proton collisions, focusing on the role of the WW contribution and its relevance w.r.t.~the LO approximation.
In particular, we will discuss in some detail HERMES kinematics, for which transverse SSA data are available. We will then give predictions for experiments at JLab with the upgrade at 12 GeV, for COMPASS at CERN, and for a future Electron-Ion Collider. In this last case we will also show some estimates for inclusive jet production.

Before presenting our results, it is worth giving some comments on the adopted kinematical configuration w.r.t.~usual experimental setups.

According to the HERMES analysis~\cite{Airapetian:2013bim}, for instance, the lepton is assumed to move along the positive $Z_{\rm cm}$ axis, so that we should consider the processes $\ell \, \pup \to h \, X$, rather than $\pup \ell \to h \, X$. In this reference frame the $\uparrow$ ($\downarrow$) direction is still along the $+Y_{\rm cm}$ ($-Y_{\rm cm}$) axis and, keeping the usual definition of $x_F = 2 P_L/\sqrt s$, where $P_L$ is the longitudinal momentum of the final hadron, only the sign of $x_F$ is reversed.

The azimuthal dependent cross section measured by HERMES is defined as~\cite{Airapetian:2013bim}:
\be
d\sigma = d\sigma_{UU}[1+S_T \, A_{UT}^{\sin\psi} \sin\psi] \>,
\label{sigH}
\ee
where
\be
\sin \psi = \hat{\bm{S}}_T \cdot (\hat{\bm{P}}_T \times \hat{\bm{k}})
\ee
coincides with our $\sin\phi_S$ of Eq.~(\ref{phis}), as $\bm{p}$ and $\bm{k}$ (respectively, the proton and the lepton 3-momenta) are opposite vectors
in the lepton-proton {\it c.m.} frame and one has:
\be
A_{UT}^{\sin\psi}(x_F, P_T) = A_N^{p^\uparrow \ell \to h X}(-x_F, P_T)  \>,
\label{AUT-hermes}
\ee
where $A_N^{p^\uparrow \ell \to h X}$ is the SSA that we compute here, and $A_{UT}^{\sin\psi}$ is the quantity measured by HERMES~\cite{Airapetian:2013bim}.

In the following, to keep uniform the presentation of our results, we will show our predictions adopting the HERMES setup also for JLab and COMPASS experiments. For EIC we prefer to keep the other configuration, with the proton moving along the positive $Z_{\rm cm}$ axis, since it allows to emphasize the strong analogies with the SSAs observed in $p^\uparrow p\to h\,X$ processes.

Finally, we notice that at relatively low $P_T$, around 1-2 GeV, due to the inclusion of transverse momentum effects one or more of the partonic Mandelstam variables might become smaller than a typical hadronic scale. This configuration would correspond to a situation where the propagator of the exchanged particle in the partonic scattering becomes soft. In order to avoid such a potential problem, following Ref.~\cite{DAlesio:2004eso}, we have introduced an infrared regulator mass ($\mu_0 =0.8$ GeV). We have checked that shifting the partonic Mandelstam invariants by this quantity squared or cutting them out below it gives similar results. Estimates will be shown adopting the shifting procedure.

\subsection{Unpolarized cross sections}

For the computation of the unpolarized cross sections within the adopted TMD approach we will use the following factorized expressions for the unpolarized TMDs:
\bea
\label{TMDs}
f_{a/p}(x,k_\perp) =  f_{a/p}(x)\, \frac{1}{\pi \langle k_\perp^2\rangle} e^{-k_\perp^2/\langle k_\perp^2\rangle} &&
D_{h/c}(z,p_\perp)  =  D_{h/c}(z)\, \frac{1}{\pi \langle p_\perp^2\rangle} e^{-p_\perp^2/\langle p_\perp^2\rangle}\,,
\eea
with $\langle k_\perp^2\rangle=0.25$ GeV$^2$ and $\langle p_\perp^2\rangle=0.2$ GeV$^2$ as extracted in Ref.~\cite{Anselmino:2005nn}.
For the collinear parton distributions, $f_{a/p}(x)$, we adopt the GRV98 set~\cite{Gluck:1998xa}, while for the collinear fragmentation functions (FFs), $D_{h/c}(z)$, we use the Kretzer set~\cite{Kretzer:2000yf} and the one by de Florian, Sassot and Stratmann (DSS)~\cite{deFlorian:2007aj}. The reasons for this choice are the following: these sets were adopted in the extraction of the Sivers and Collins functions we use here for the calculation of the SSAs (next Section); they are characterized by a different role of the gluon fragmentation function, that could play a role in the WW contribution.

\subsubsection{HERMES}
\label{unp-hermes}

\begin{figure}[ht!]
 \begin{center}
 \includegraphics[scale=.9]{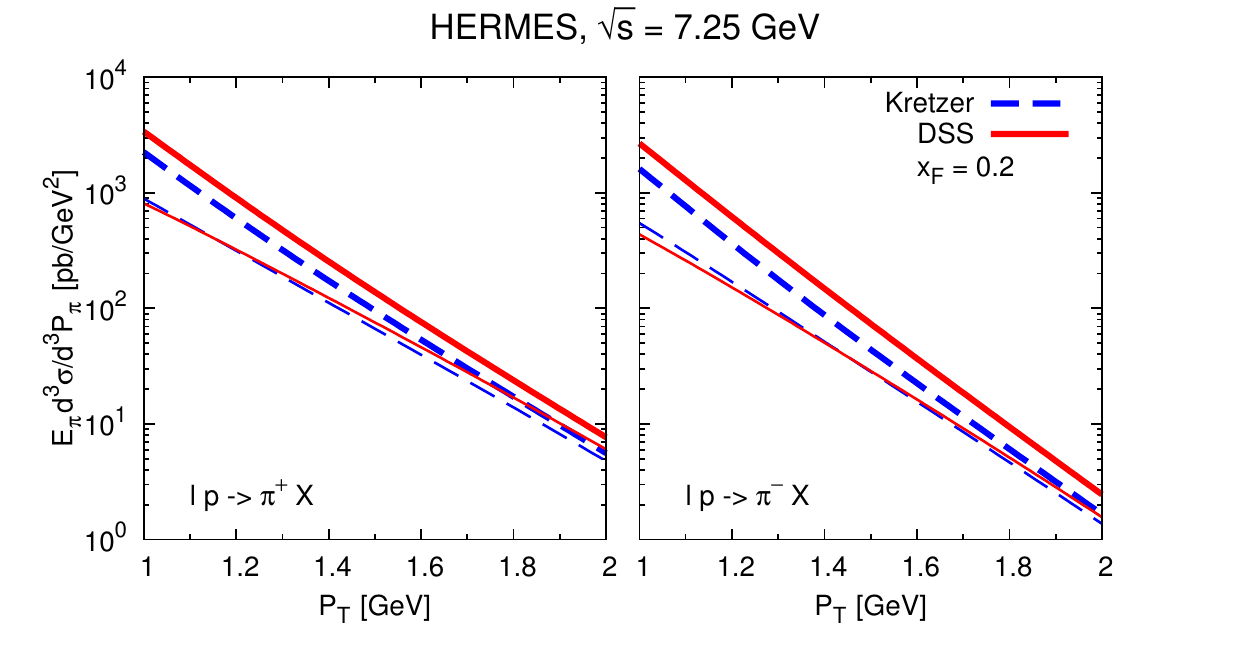}
\end{center}
 \caption{Estimates of the unpolarized cross sections at $x_F=0.2$ as a function of $P_T$ for $\pi^+$ (left panel) and $\pi^-$ (right panel) production in $\ell\, p\to \pi\,X$, at HERMES, $\sqrt{s} = 7.25$ GeV, adopting two sets for the fragmentation functions: Kretzer set (blue dashed lines) and DSS set (red solid lines). The thin curves represent the LO calculation, while the thick curves the total (LO+WW) result.}
  \label{fig:unp-herm-xf02}
\end{figure}

\begin{figure}[ht!]
 \begin{center}
 \includegraphics[scale=.9]{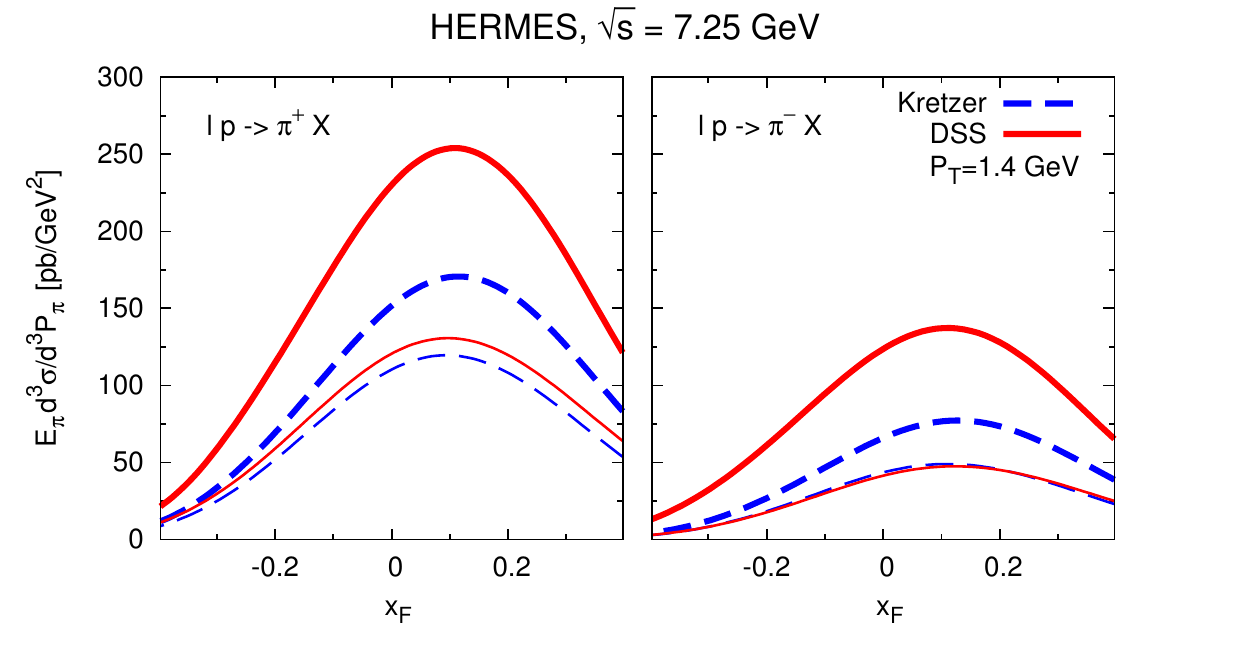}
\end{center}
 \caption{Estimates of the unpolarized cross sections at $P_T=1.4$ GeV as a function of $x_F$ for $\pi^+$ (left panel) and $\pi^-$ (right panel) production in $\ell\, p\to \pi\,X$, at HERMES, $\sqrt{s} = 7.25$ GeV. Curves have the same meaning as in Fig.~\ref{fig:unp-herm-xf02}.}
  \label{fig:unp-herm-pt14}
\end{figure}

In Figs.~\ref{fig:unp-herm-xf02} and \ref{fig:unp-herm-pt14} we present our estimates for the unpolarized cross sections for $\pi^+$ (left panels) and $\pi^-$ (right panels) production at $\sqrt s\simeq 7.25$ GeV, respectively at fixed $x_F=0.2$ as a function of $P_T$, and at fixed $P_T=1.4$~GeV as a function of $x_F$. The thin curves refer to the LO calculation, while the thick ones to the total (LO+WW) contribution. In particular, the blue dashed lines are obtained adopting the Kretzer set for the fragmentation functions, while the red solid lines with the DSS set.

We start noticing that at LO there are almost no differences between the estimates based on the two FF sets, while these become more significant when also the WW piece is included. The reason is due to the much larger gluon fragmentation function in the DSS set w.r.t.~the Kretzer one, that enters through the $\gamma q\to g q$ process. More interesting from our point of view are the following features: from Fig.~\ref{fig:unp-herm-xf02} we see that the WW piece plays a more relevant role at smaller $P_T$, being almost three times bigger than the LO term around $P_T=1$ GeV. This can be ascribed to the smaller values of $y$ reached at low $P_T$ and the corresponding enhancing factor coming from the WW distribution (see Eq.~(\ref{ww})). Moreover, as one can see in Fig.~\ref{fig:unp-herm-pt14}, its contribution is strongly asymmetric in $x_F$ (more than the LO term), being more important for (large) positive $x_F$ values of the final hadron. This could appear surprising, since in such a configuration the lepton undergoes, on average, a backward scattering, and one would expect a lesser role from quasireal photon exchange. On the other hand for large positive $x_F$, when the final hadron (as well as its parent parton $c$) is produced in the \emph{backward} proton hemisphere, $|\hat u| \ll |\hat t|$, where $\hat t = (p_a - p_c)^2$ and $\hat u = (p_\gamma-p_c)^2$ for the $a\gamma\to c d$ process. This is the region favoured by the WW contribution w.r.t.~the LO piece since this one goes like $1/Q^2 \equiv 1/\hat t^2$, while the partonic cross section for the dominant subprocess $q\gamma\to q g$ (see Eqs.~(\ref{updownqgqg2}) and (\ref{Mqgaqg})) goes like $1/\hat s\hat u$.

\subsubsection{Jefferson Lab at 12 GeV}

We consider the process $\ell\,^3$He $\to \pi X$ and, adopting $SU(2)$ symmetry, give estimates at $\sqrt s= 4.84$ GeV for the cross section \emph{per nucleon}. In particular we plot
\be
\frac{d^2\sigma}{dx_FdP_T} = \frac{2\pi P_T}{\sqrt{x_F^2+ x_T^2}} \, E_\pi \frac{d^3\sigma}{d^3\bm{P}_\pi}\,
\ee
where $x_T= 2 P_T/\sqrt s$.

\begin{figure}[ht!]
 \centering
 \includegraphics[scale=0.9]{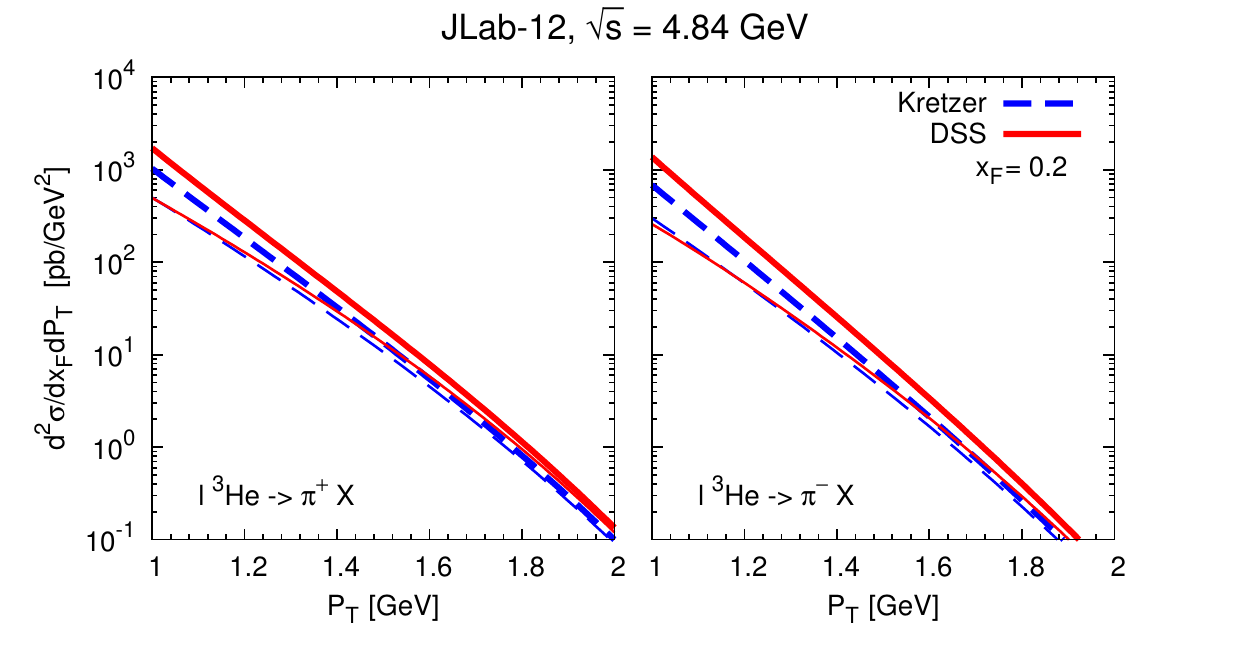}
 \caption{Estimates of the unpolarized cross sections \emph{per nucleon} at $x_F=0.2$ as a function of $P_T$ for $\pi^+$ (left panel) and $\pi^-$ (right panel) production in $\ell\,^3$He $\to \pi\,X$, at JLab-12, $\sqrt{s} = 4.84$ GeV. Curves have the same meaning as in the previous figures.
}
  \label{fig:unp-jlab-xf02}
\end{figure}

\begin{figure}[ht!]
 \centering
 \includegraphics[scale=0.9]{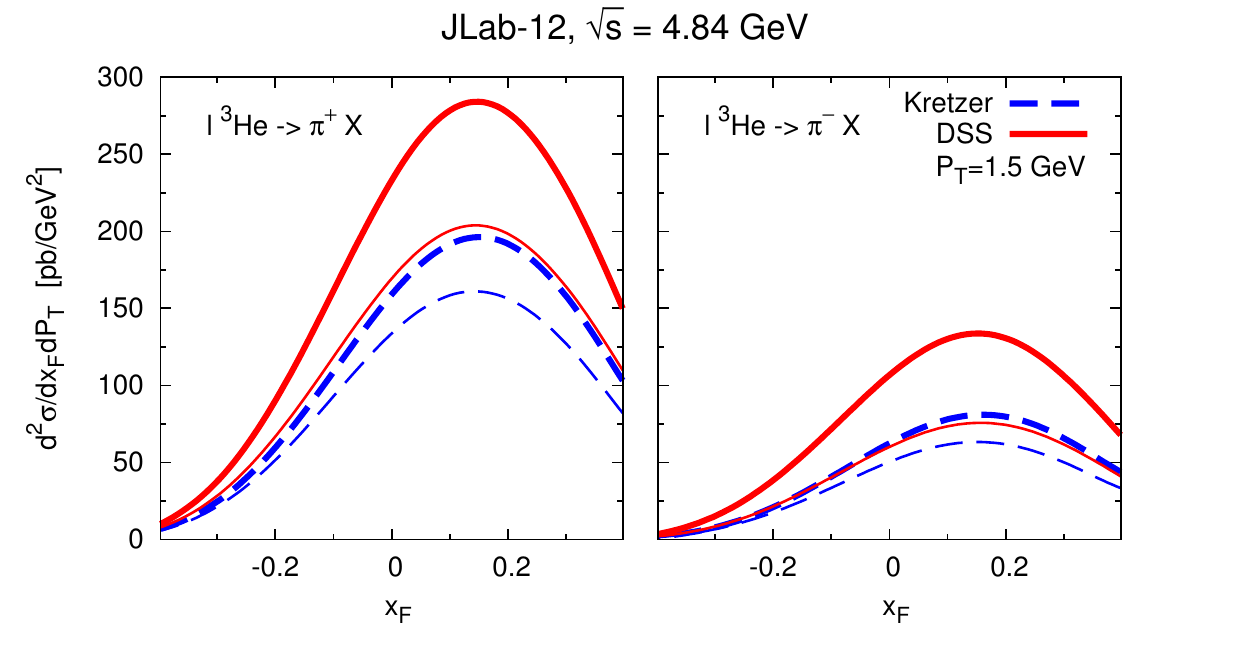}
 \caption{Estimates of the unpolarized cross sections \emph{per nucleon} at $P_T=1.5$ GeV as a function of $x_F$ for $\pi^+$ (left panel) and $\pi^-$ (right panel) production in $\ell \,^3$He $\to \pi\,X$, at JLab-12, $\sqrt{s} = 4.84$ GeV. Curves have the same meaning as in the previous figures.}
  \label{fig:unp-jlab-pt15}
\end{figure}

In Figs.~\ref{fig:unp-jlab-xf02} and \ref{fig:unp-jlab-pt15} we present the unpolarized cross sections for $\pi^+$ (left panels) and $\pi^-$ (right panels) production at $\sqrt s\simeq 4.84$ GeV, respectively at fixed $x_F=0.2$ as a function of $P_T$, and at fixed $P_T=1.5$ GeV as a function of $x_F$. The curves have the same meaning as for the HERMES kinematics. Same considerations are also valid, with the only extra remark, see Fig.~\ref{fig:unp-jlab-pt15}, that even the LO calculation gives sizeably different results adopting the two FF sets. This is due to the more important role of the DSS FFs in the very large-$z$ region, as explored at this energy.

\subsubsection {COMPASS}

\begin{figure}[ht!]
 \centering
 \includegraphics[scale=0.9]{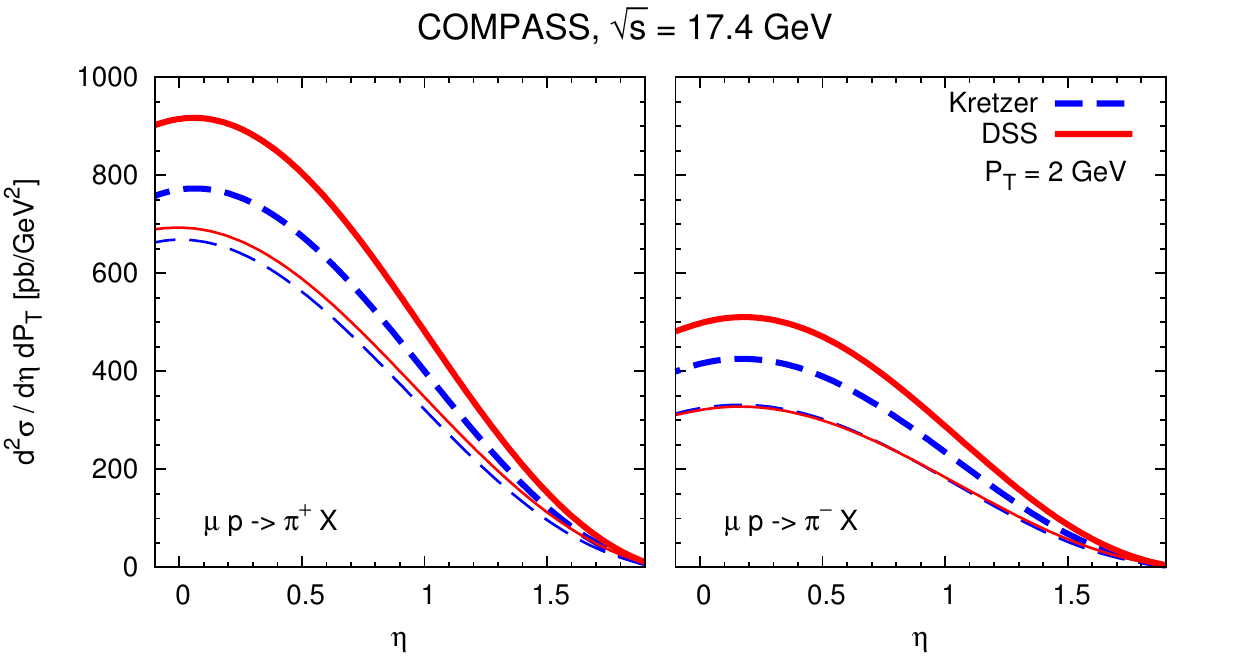}
 \caption{Estimates of the unpolarized cross sections at $P_T=2$ GeV as a function of $\eta$ for $\pi^+$ (left panel) and $\pi^-$ (right panel) production in $\mu \, p \to \pi\,X$, at COMPASS, $\sqrt{s} = 17.4$ GeV. Curves have the same meaning as in the previous figures.
}
  \label{fig:unp-comp-pt2}
\end{figure}

For the COMPASS experiment the incoming lepton is a muon with a lab-energy of 160 GeV, resulting in $\sqrt s = 17.4$ GeV. Following their setup, we use the \emph{c.m}.~pseudorapidity $\eta$ of the produced hadron in the range $-0.1<\eta < 2$ (as covered by COMPASS spectrometer). Similarly to the HERMES configuration, pseudorapidity is counted as positive in the forward direction of the incident muon. We have
\be
\frac{d^2\sigma}{d\eta dP_T} = 2\pi P_T\, E_\pi \frac{d^3\sigma}{d^3\bm{P}_\pi}\,.
\ee
In Fig.~\ref{fig:unp-comp-pt2} we show the unpolarized cross sections for $\pi^+$ (left panel) and $\pi^-$ (right panel) production at $\sqrt s = 17.4$~GeV and fixed $P_T=2$ GeV as a function of $\eta$. The curves have the same meaning as in the previous figures. In this case the two FF sets give almost the same LO results. At variance with what happens for the HERMES configuration, here the WW contribution, although still relevant, plays a lesser role: adopting the DSS set, for instance, it is at most 65\% of the LO term for $\pi^-$ and only 30\% for $\pi^+$ production. Indeed, the muon mass is almost 200 times bigger than the electron mass, thus reducing the size of the logarithmic piece entering Eq.~(\ref{ww}), partially cancelled by the finite term.

\subsubsection{Electron-Ion Collider}

The proposed future EIC with $\sqrt s = 100$ GeV~\cite{Accardi:2012qut} will allow to study the cross section for single-inclusive pion production in electron-proton collisions at very high energies, comparable to those reached in  proton-proton reactions. In such a case we prefer to adopt the configuration in which the proton moves along the positive $Z_{\rm cm}$ axis, defining $x_F$ accordingly (more precisely, $x_F>0$ here will refer to the \emph{forward} proton hemisphere). This choice will appear more natural and helpful in the context of the analysis of transverse single-spin asymmetries (next Section), allowing an easier comparison with $A_N$ measured in $pp$ collisions. For the same reason we will consider neutral pion production.

\begin{figure}[ht!]
 \centering
 \includegraphics[scale=0.9]{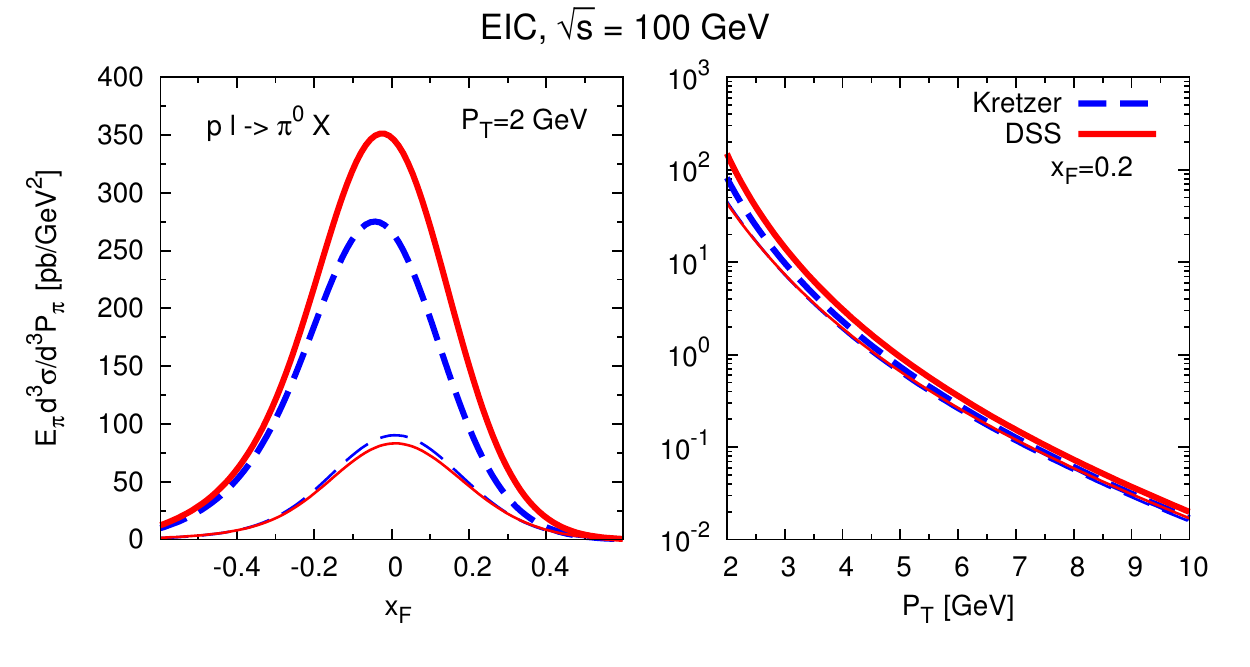}
 \caption{Estimates of the unpolarized cross sections in $p \,\ell \to \pi^0\,X$, at EIC, $\sqrt{s} = 100$ GeV, at $P_T=2$ GeV as a function of $x_F$ (left panel) and at fixed $x_F=0.2$ as a function of $P_T$ (right panel). Notice that $x_F>0$ here corresponds to the forward proton hemisphere. Curves have the same meaning as in the previous figures.
}
  \label{fig:unp-eic-pi0}
\end{figure}

In Fig.~\ref{fig:unp-eic-pi0} we show the unpolarized cross sections for $\pi^0$ production at $\sqrt s = 100$~GeV, respectively at fixed $P_T=2$ GeV as a function of $x_F$ (left panel) and at fixed $x_F=0.2$ as a function of $P_T$ (right panel). The curves have the same meaning as in the previous figures. Once again the two FF sets give almost the same LO results. At variance with what happens at lower energies, here the WW contribution turns out to be much more relevant, being up to four times larger than the LO term at $P_T=2$ GeV. The reason is that at such large energies and not so large $P_T$, for $x_F\geq 0$ we probe the small-$y$ region of the photon spectrum in the WW distribution, that behaves like $1/y$, while for $x_F <0$ (the backward region here)  the WW partonic cross sections dominate the LO one,  since $|\hat u| \ll |\hat t|$.

\begin{figure}[ht!]
 \centering
 \includegraphics[scale=0.9]{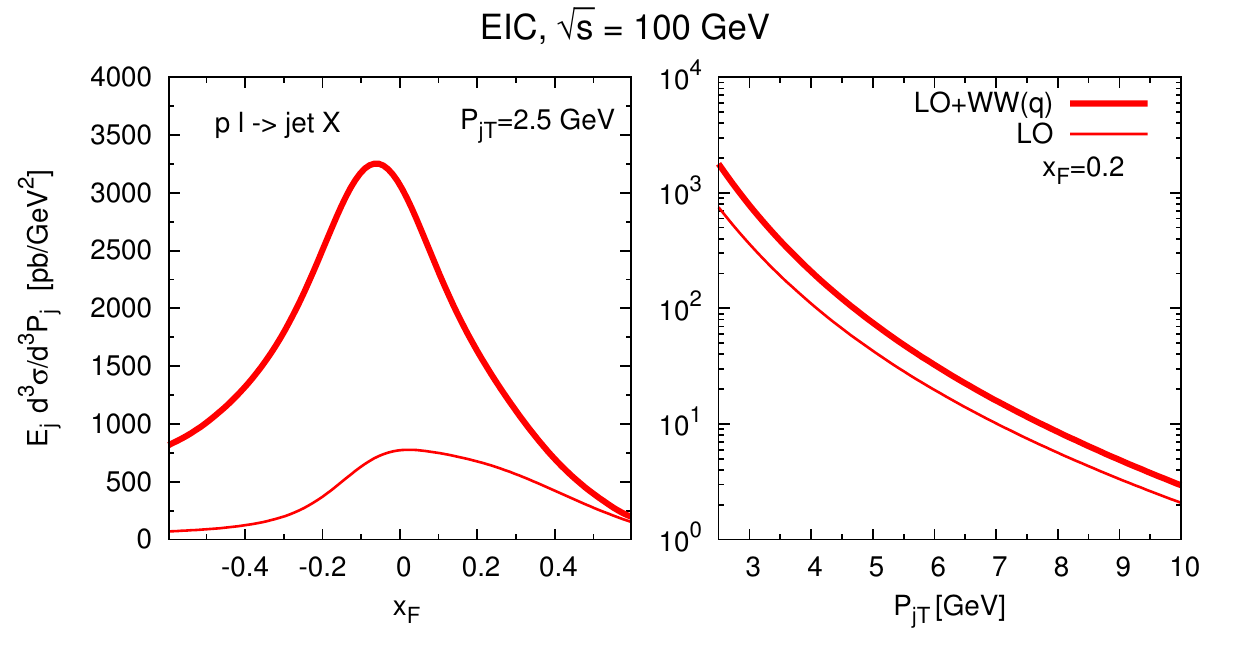}
 \caption{Estimates of the unpolarized cross sections in $p\, \ell \to {\rm jet} \,X$, at EIC, $\sqrt{s} = 100$ GeV at $P_{{\rm j}T}=2.5$ GeV as a function of $x_F$ (left panel) and at $x_F=0.2$ as a function of $P_{{\rm j}T}$ (right panel). Notice that $x_F>0$ here corresponds to the forward proton hemisphere. Thick (thin) curves refer  to the LO+WW (LO) contribution.
}
  \label{fig:unp-eic-jet}
\end{figure}

At EIC, given the large energy available, the interesting study of inclusive jet production could be feasible. In Fig.~\ref{fig:unp-eic-jet} we give some estimates of the cross sections for jet production at fixed $P_{{\rm j}T}=2.5$ GeV as a function of $x_F = 2 P_{\rm{j} L}/\sqrt s$ (left panel) and at fixed $x_F=0.2$ as a function of $P_{{\rm j}T}$ (right panel).
The slightly larger $P_{{\rm j}T}$ value considered helps keeps potential infrared divergences in the hard elementary scattering under better control. Even here the WW contribution heavily dominates the LO term over almost the full $x_F$ range (with its characteristic asymmetric behaviour). Notice that in this case both at large positive and large negative $x_F$ there is no dilution from the large-$z$ behaviour of fragmentation functions as happens in inclusive pion production.

\subsection{Transverse single-spin asymmetries}

We now focus on the main issue of this paper, the study of the role of quasireal photon exchange in SSAs for single-inclusive particle production in lepton-proton collisions, starting with a comparison with the available data from the HERMES Collaboration~\cite{Airapetian:2013bim}.
In our computations, based on a TMD factorization scheme, we consider two different sets of the quark Sivers and Collins functions (the latter coupled to the transversity distribution), as previously obtained in a series of papers from fits of SIDIS and $e^+e^-$
data~\cite{Anselmino:2005ea,Anselmino:2007fs,Anselmino:2008sga,Anselmino:2008jk}.

These sets, besides some differences in the initial assumptions and in the data used for their extraction, differ in the choice of the collinear fragmentation functions. More precisely, for the fits~\cite{Anselmino:2005ea} and~\cite{Anselmino:2007fs} (SIDIS~1) the Kretzer FF set was adopted, while for the fits~\cite{Anselmino:2008sga} and ~\cite{Anselmino:2008jk} (SIDIS~2) the DSS FF set was employed. The SIDIS~1 and SIDIS~2 sets are well representative of the extractions of these TMDs and their uncertainties. Concerning the gluon Sivers function, still poorly determined, we adopt the recent extractions of Ref.~\cite{D'Alesio:2015uta}. Notice that these are obtained assuming a specific set for the quark Sivers functions and therefore we will have a gluon Sivers function associated to each SIDIS set. It is worth recalling that the extractions of the quark and gluon Sivers functions (as well of the transversity distribution) are constrained only up to $x\simeq 0.3$.

In the following we will consider both the fully-inclusive HERMES data, already discussed in Ref.~\cite{Anselmino:2014eza}, as well as the sub-sample of anti-tagged data (with no detection of the final lepton), for $\ell \, p^\uparrow \to \pi \, X$ processes at large $P_T$. In both cases there is only one large scale (needed for a perturbative calculation), the $P_T$ of the final pion. For this reason we only look at those data at $P_T\ge$ 1 GeV.

At variance with SIDIS azimuthal asymmetries, where the single contributions to $A_N$ coming from the Sivers and Collins effects can be accessed separately by looking at their proper azimuthal modulations, here the two effects could contribute together and mix up. For this reason we will present for each SIDIS set the overall contribution, adding together the quark Sivers (dominant) and Collins (almost negligible) effects. This will be done for the LO and the complete (LO+WW) calculation. For this last one we will also show the overall statistical uncertainty bands given as the envelope of the uncertainties on the quark Sivers and Collins functions, obtained following the procedure described in Appendix A of Ref.~\cite{Anselmino:2008sga}. For completeness, but with a word of caution, we have also computed the results obtained adding the contribution from the gluon Sivers function. In the following we will show them explicitly only for HERMES kinematics.

\subsubsection{HERMES: SSAs and comparison with data}

\begin{figure}[ht!]
 \centering
 \includegraphics[scale=0.9]{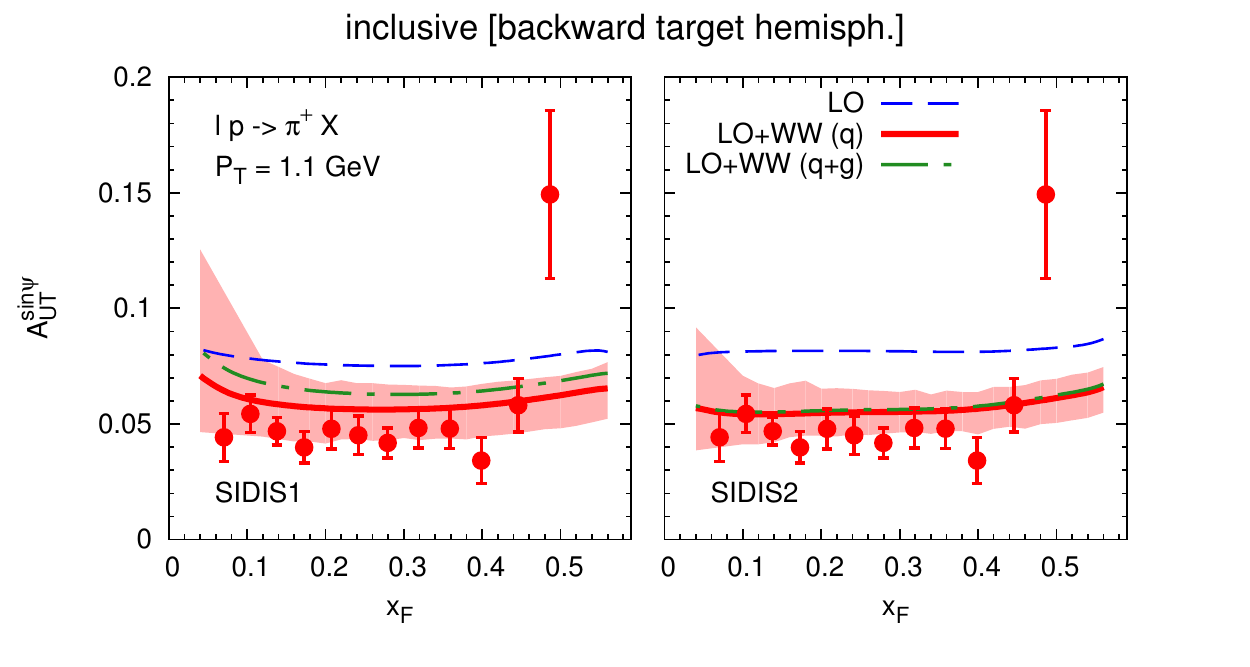}
 \includegraphics[scale=0.9]{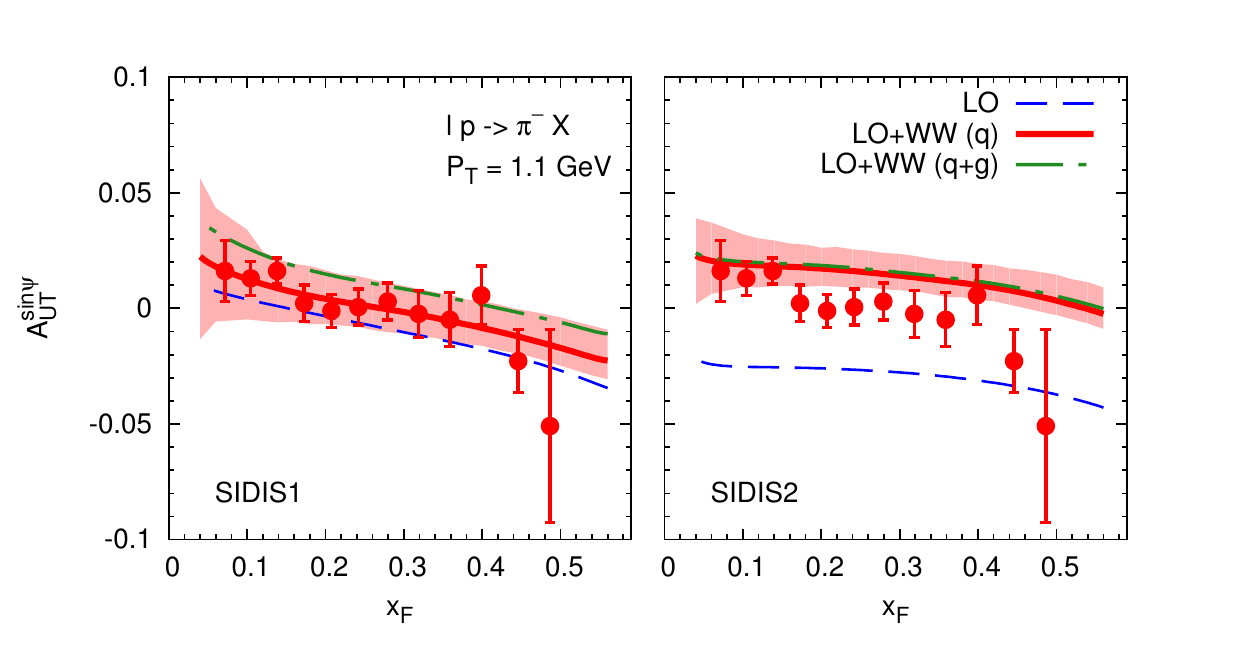}
 \caption{Theoretical estimates for $A_{UT}^{\sin\psi}$ vs.~$x_F$ at $\sqrt{s}\simeq 7.25$ GeV and $P_T = 1.1$ GeV for inclusive $\pi^+$ (upper panels) and $\pi^-$ (lower panels) production in $\ell \, \pup \to \pi \, X$ processes, compared with the fully-inclusive HERMES data~\cite{Airapetian:2013bim}. Two sets for the Sivers and Collins functions have been considered: the SIDIS 1 set (left panels) and the SIDIS 2 set (right panels). More precisely, we show both the LO (blue dashed lines) and LO+WW (red solid lines) quark contributions, as well as the total result including the gluon Sivers effect (green dot-dashed lines). The overall statistical uncertainty band, obtained following the procedure described in Appendix A of Ref.~\cite{Anselmino:2008sga} is also shown.
}
  \label{fig:SSA-pi-pt11}
\end{figure}

Our predictions for $A_{UT}^{\sin\psi}$, for inclusive $\pi^+$ (upper panels) and $\pi^-$ (lower panels) production, as a function of $x_F$ at $P_T = 1.1$ GeV, compared with the fully-inclusive HERMES data~\cite{Airapetian:2013bim}, are presented in Fig.~\ref{fig:SSA-pi-pt11} (this is the only bin at relatively large $P_T$). More precisely, we show the LO calculation, blue dashed lines, and the complete result adding the WW piece, red solid lines, adopting the quark Sivers and Collins functions from the SIDIS 1 (left panels) and SIDIS 2 (right panels) sets. The overall statistical uncertainty band is also shown. The green dot-dashed lines represent the total contribution including also the gluon Sivers effect.

We can then make the following remarks: the inclusion of the WW contribution (that in this kinematical region dominates the unpolarized cross sections) improves significantly the agreement with the data; the Collins effect is always tiny or completely negligible (both in the LO and WW contributions); the differences between the predictions adopting the SIDIS~1 and SIDIS~2 sets are due to the different behaviour of the corresponding Sivers functions; the contribution coming from the gluon Sivers function is almost negligible for the SIDIS~2 set, while that for the SIDIS~1 set is relatively more important, reducing the agreement with the data. We have nevertheless to point out that there is still a large uncertainty in the gluon Sivers function extraction in the large-$x$ region, as covered in such a kinematical configuration.

\begin{figure}[ht!]
 \centering
 \includegraphics[scale=0.9]{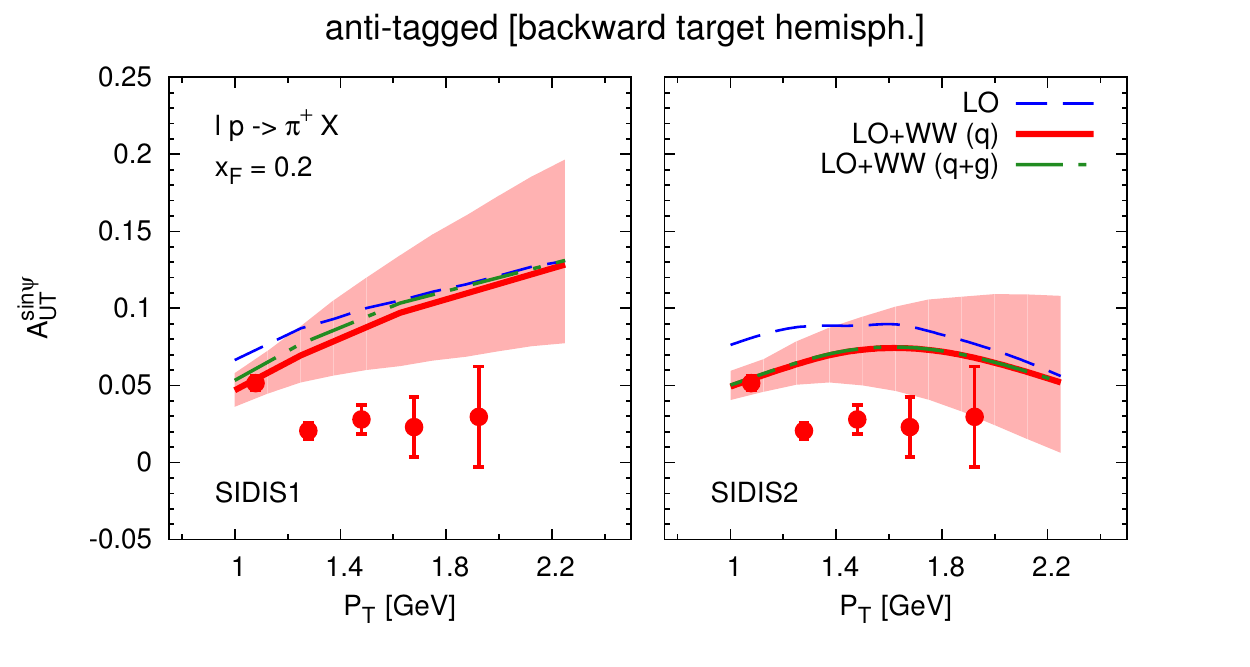}
 \includegraphics[scale=0.9]{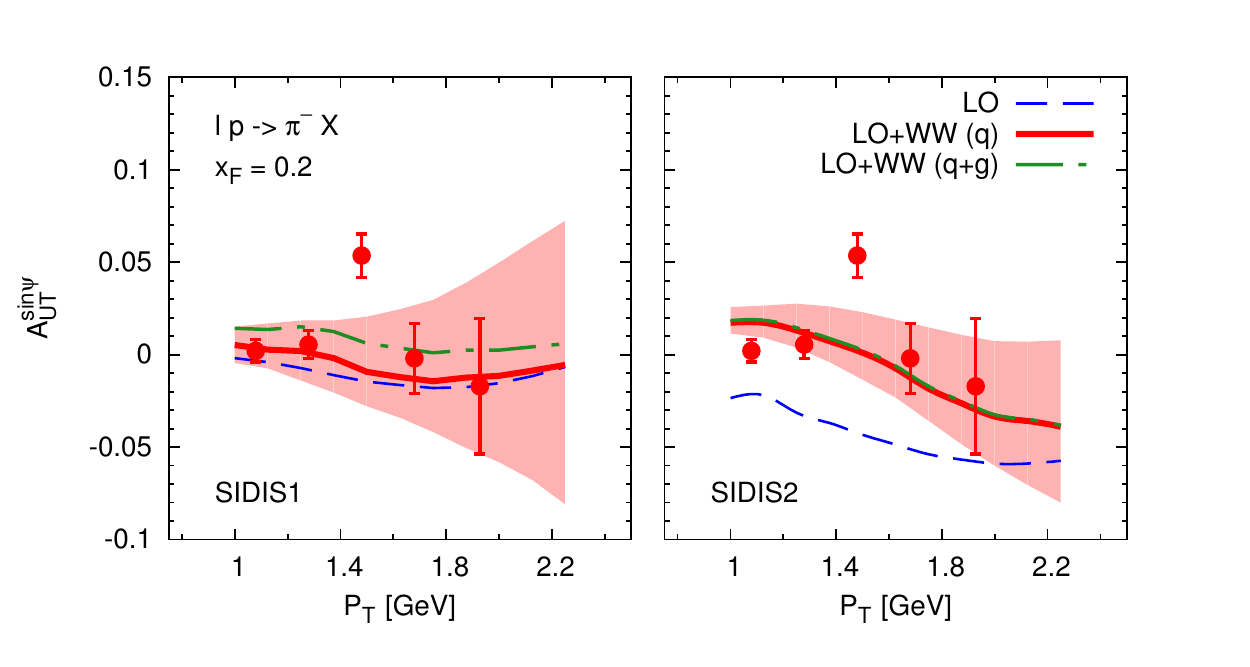}
 \caption{Theoretical estimates for $A_{UT}^{\sin\psi}$ vs.~$P_T$ at $\sqrt{s}\simeq 7.25$ GeV and $x_F = 0.2$ for inclusive $\pi^+$ (upper panels) and $\pi^-$ (lower panels) production in $\ell \, \pup \to \pi \, X$ processes, compared with the anti-tagged HERMES data~\cite{Airapetian:2013bim}. Curves have the same meaning as in the previous figure.
}
  \label{fig:SSA-pi-xf02}
\end{figure}

In Fig.~\ref{fig:SSA-pi-xf02} we present, for the first time, our results for the anti-tagged category for $A_{UT}^{\sin\psi}$, compared with HERMES data~\cite{Airapetian:2013bim}, at fixed $x_F=0.2$ (average value of the data set) as a function of $P_T$. Once again we consider the inclusive $\pi^+$ (upper panels) and $\pi^-$ (lower panels) production, adopting two sets for the quark Sivers and Collins functions: SIDIS 1 (left panels) and SIDIS 2 (right panels). The curves have the same meaning as for the fully-inclusive case. From these results we can observe that once again the WW contribution leads to a much better description of the data (even if some sizeable discrepancy for the $\pi^+$ data remains). The gluon Sivers effect is negligible, except for the SIDIS~1 set in $\pi^-$ production. However, this kinematical region probes the still poorly constrained large-$x$ behaviour of the Sivers functions (the dominant contribution), which reflects into wider statistical error bands.

\subsubsection{SSAs at JLab}

Forthcoming measurements at the energy of 12 GeV are going to be performed at JLab (the 6 GeV energy setup is not able to provide sufficiently large $P_T$ values), on transversely polarized proton, neutron and deuteron targets. We focus here, for its complementarity with HERMES data, on neutron target. Indeed, the combined analysis of proton and neutron
target events will help in our understanding of the flavour decomposition and on the role of the up and down quark contributions.

\begin{figure}[ht!]
 \centering
 \includegraphics[scale=0.9]{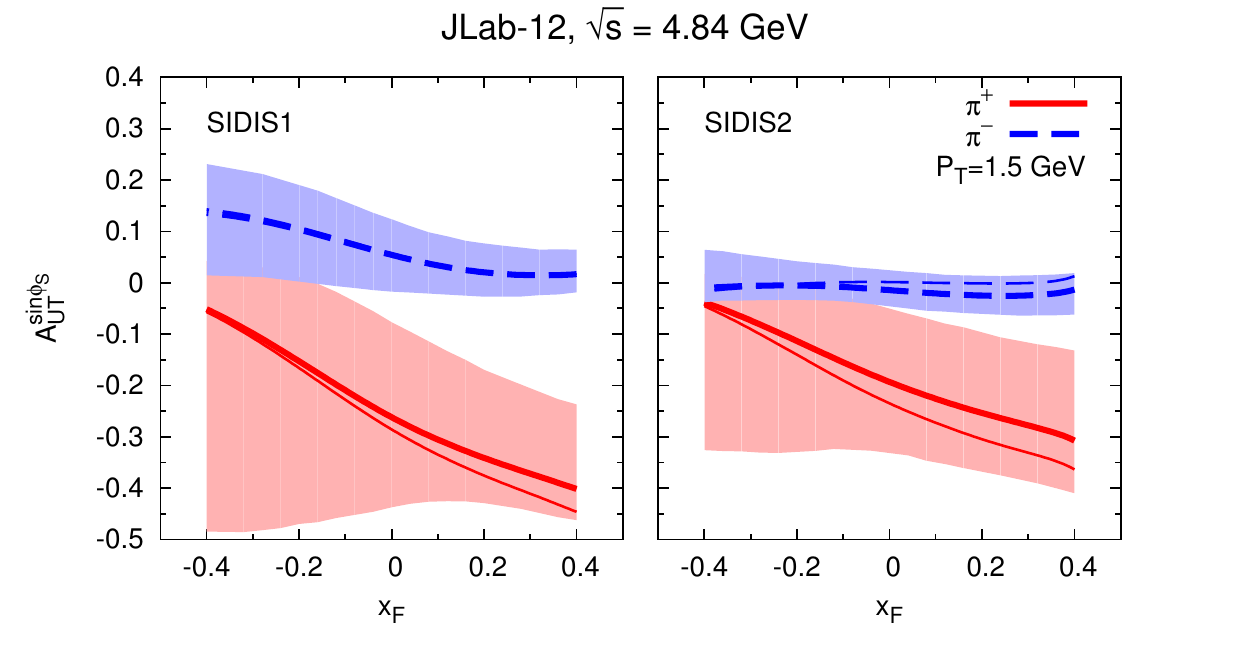}
 \caption{Theoretical estimates for $A_{UT}^{\sin\phi_S}$ vs.~$x_F$ at $\sqrt{s}\simeq 4.84$ GeV and $P_T = 1.5$ GeV for inclusive $\pi^+$ (red solid lines) and $\pi^-$ (blue dashed lines) production, which will be measured at JLab operating on a polarized $^3$He (neutron) target, with a beam energy of 12 GeV. The thin curves refer to the LO calculation, while the thick ones to the full, LO+WW, estimates for the two sets for the quark Sivers and Collins functions: SIDIS~1 (left panel) and SIDIS~2 (right panel). The overall statistical uncertainty band, obtained following the procedure described in Appendix~A of Ref.~\cite{Anselmino:2008sga}, is also shown.
}
  \label{fig:SSA-jlab-pi-pt15}
\end{figure}

Our estimates for the JLab SSAs, $A_{UT}^{\sin\phi_S}$, for inclusive pion production off polarized $^3$He (neutron) target are shown in Fig.~\ref{fig:SSA-jlab-pi-pt15} at fixed $P_T=1.5$ GeV as a function of $x_F$. Notice that this somehow large $P_T$ value has been chosen for uniformity with what discussed for the unpolarized cross section and because it allows to span a larger region in $x_F$ (in particular its positive values). We recall that we keep adopting the HERMES configuration, with the incoming lepton moving along the positive $Z_{\rm cm}$ axis and plot $A_{UT}^{\sin\phi_S}\equiv A_{UT}^{\sin\psi}$. In particular, we show for the SIDIS~1 (left panel) and the SIDIS~2 (right panel) the LO (thin lines) and the LO+WW (thick lines) calculation, displaying also the uncertainty bands for the total contribution. The gluon Sivers effect, not included, plays some role only in $\pi^-$ production when adopting the SIDIS~1 set, as in the HERMES case, being otherwise negligible. In most cases the full, LO+WW, estimates present the same behaviour, in size and sign, as the LO ones. Nevertheless one has to keep in mind that, with the DSS FF set for instance, the WW piece alone is about 50\% (90\%) of the LO contribution for $\pi^+$ ($\pi^-$) production in this kinematical region.
The wider uncertainty bands are due to the large-$x$ region probed at such moderate energies, where the current extractions of the Sivers functions are still unconstrained.

It is worth noticing that the differences with respect to the corresponding HERMES results are due to the exchanged role of the up and down quark Sivers distributions when adopting SU(2) symmetry for a neutron target (JLab). This reflects also in the relative weight of the WW piece when going from positively to negatively charged pions. Moreover, the very large size of $A_N$ for $\pi^+$ production at large positive $x_F$ (backward neutron hemisphere), both adopting the SIDIS~1 and the SIDIS~2 sets, is due to the probed values of the quark light-cone momentum fraction in the polarized neutron (down to 0.1 at $x_F>0$). For such $x$ values the negative up quark neutron Sivers function (down quark for a proton) is quite large for these two sets and, coupling to the favored fragmentation function, largely dominates over the other contributions.

\subsubsection{SSAs at COMPASS}

\begin{figure}[ht!]
 \centering
 \includegraphics[scale=0.9]{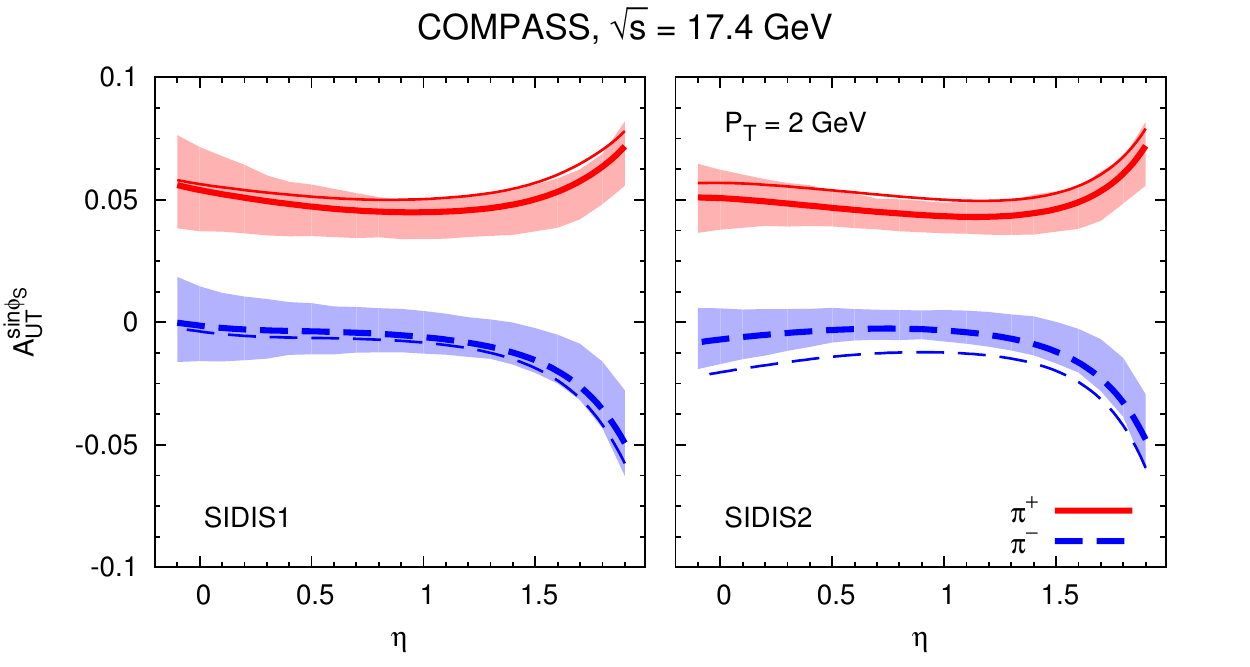}
 \caption{Theoretical estimates for $A_{UT}^{\sin\phi_S}$ vs.~$x_F$ at $\sqrt{s}\simeq 17.4$ GeV and $P_T = 2$ GeV for inclusive $\pi^+$ (red solid lines) and $\pi^-$ (blue dashed lines) production in $\mu p^\uparrow\to \pi\, X$ at COMPASS. Two sets for the quark Sivers and Collins functions have been adopted: SIDIS 1 (left panel) and SIDIS 2 (right panel). The overall statistical uncertainty band are also shown. Curves have the same meaning as in the previous figure.
}
  \label{fig:SSA-comp-pi-pt2}
\end{figure}

Another place where these SSAs could be measured is certainly the COMPASS experiment. Here we present some estimates for this experimental setup. In Fig.~\ref{fig:SSA-comp-pi-pt2} we show $A_{UT}^{\sin\phi_S}$ vs.~$x_F$ at $\sqrt{s}\simeq 17.4$ GeV and $P_T = 2$ GeV for inclusive $\pi^+$ (red solid lines) and $\pi^-$ (blue dashed lines) production in $\mu \, p^\uparrow\to \pi\, X$. Curves have the same meaning as in the previous figures. One can see that the SSAs for $\pi^+$ production are expected to be sizeable, with quite narrow error bands: a clear test of this approach could be then carried out. Again the inclusion of the WW contribution changes only slightly the LO estimates.

\subsubsection{SSAs at EIC}

In Refs.~\cite{Anselmino:2009pn,Anselmino:2014eza} some estimates for inclusive jet and inclusive neutral pion production for an Electron-Nucleon Collider at 50 GeV were given with the aim of checking whether some features of the SSAs observed in $p^\uparrow p\to\pi\,X$, and reproduced in a TMD scheme, could be also encountered in the process under consideration. In such a case it is more convenient to adopt the configuration where the polarized proton is moving along the positive $Z_{\rm cm}$ axis and positive $x_F$ values correspond to the forward proton hemisphere.

\begin{figure}[ht!]
 \centering
 \includegraphics[scale=0.9]{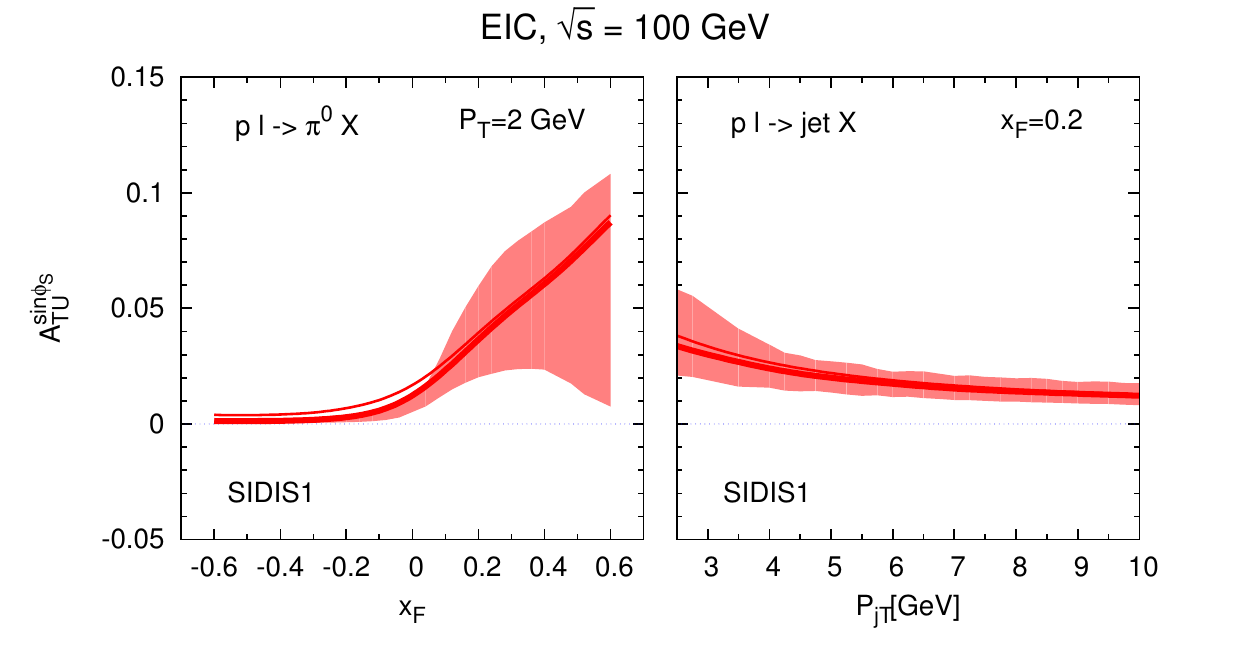}
 \caption{Theoretical estimates for $A_{TU}^{\sin\phi_S}$ at $\sqrt{s}= 100$ GeV as a function of $x_F$ at $P_T = 2$~GeV for inclusive $\pi^0$ production (left panel) and as a function of $P_{{\rm j}T}$ at $x_F=0.2$  for inclusive jet production (right panel), adopting the SIDIS 1 set for the quark Sivers and Collins functions. The overall statistical uncertainty bands are also shown. Thick (thin) lines refer to the LO+WW (LO) calculation.
}
  \label{fig:SSA-eic-pi}
\end{figure}

It is then interesting to see what happens when one includes also the contribution from quasireal photon exchange at the future EIC.

In Fig.~\ref{fig:SSA-eic-pi} we show our estimates at $\sqrt{s}= 100$  GeV for $A_{TU}^{\sin\phi_S} \equiv A_N$ (as defined in Eq.~(\ref{ATU})) for inclusive $\pi^0$ production in $p^\uparrow \ell \to \pi\, X$ vs.~$x_F$ at $P_T = 2$ GeV (left panel) and for inclusive jet production vs.~$P_{{\rm j}T}$ at $x_F=0.2$ (right panel), adopting the SIDIS~1 set. This set indeed is the one that better reproduces the behaviour of $A_N$ in $\pup p\to\pi\,X$ processes (see for instance Ref.~\cite{Boglione:2007dm}) and that is consistent with the findings of a dedicated study performed in Ref.~\cite{Anselmino:2013rya}. Again thick (thin) curves represent LO+WW (LO) contributions. Some comments are in order:
\begin{itemize}
\item The gluon Sivers effect (not shown) is completely negligible.
\item The corresponding results for $\pi^0$ as a function of $P_T$, not shown, are almost identical to those for inclusive jet production. The same is true for $A_N$ for inclusive jet production as a function of $x_F$, not shown, almost identical to that for $\pi^0$ production.
\item As one can see the WW contribution does not change the LO behaviour. This could be expected since both contributions enter with the same structure in the SSA. We then confirm all findings of Ref.~\cite{Anselmino:2014eza} concerning the $x_F$ behaviour, with the extra important information that at such energies and $P_T$ values the WW piece is the dominant one in the unpolarized cross sections.
\item Quite interesting, the $P_T$ behaviour, shown here for the first time, is almost flat, and measurable, up to very large $P_T$ values. This is strongly analogous to what happens in $p^\uparrow p\to \pi^0 \, X$ as measured by the STAR Collaboration~\cite{Igo:2012} and it would be another very important test of the full approach.
\item The large error bands at large $x_F$ are due to the still poor knowledge of the Sivers function in the large $x$ region. Future measurements at JLab could definitely help in this respect.
\end{itemize}

\subsubsection{Results from new extractions of the Sivers and Collins functions}

At the very last stage of this work, a new extraction of the Sivers functions from the latest SIDIS data has been released~\cite{Anselmino:2016uie}. Together with the fit of the Collins and the transversity functions of Ref.~\cite{Anselmino:2015sxa}, they represent the most updated information on the relevant TMDs entering the present analysis.
Among the main features of these extractions we mention: the use of the DSS FF set with different Gaussian widths for the unpolarized TMDs, as extracted from SIDIS multiplicities~\cite{Anselmino:2013lza}, $\langle k_\perp^2 \rangle = 0.57$ GeV$^2$ and $\langle p_\perp^2 \rangle = 0.12$ GeV$^2$ (to be compared with those used in SIDIS~1 and SIDIS~2 fits, see Eq.~(\ref{TMDs})); the resulting reduced size of the $x$-dependent part of the valence up and down Sivers distributions; a more flexible parametrization of the Collins functions, with a more accurate extraction of their transverse momentum dependence. We also notice that for the new fit of the Sivers functions the CTEQ6L parton distribution functions~\cite{Pumplin:2002vw} were used.

\begin{figure}[ht!]
 \centering
 \includegraphics[scale=0.9]{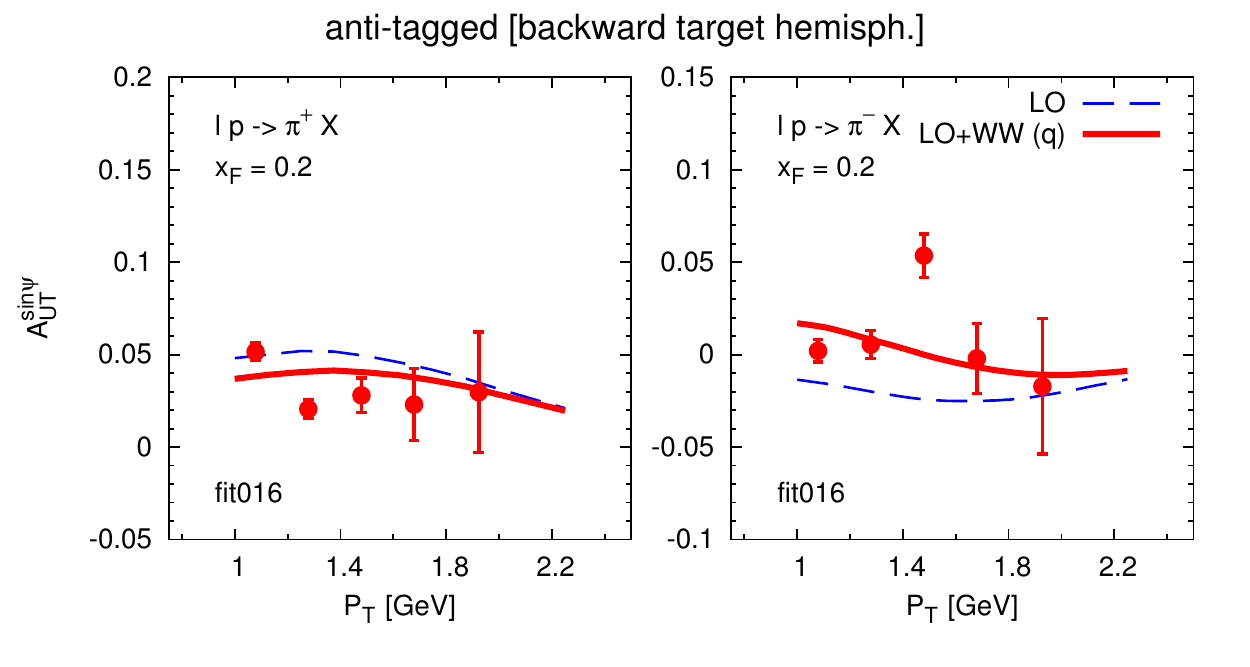}
 \caption{Theoretical estimates of the Sivers contribution to $A_{UT}^{\sin\psi}$ vs.~$P_T$ at $\sqrt{s}\simeq 7.25$ GeV and $x_F = 0.2$ for inclusive $\pi^+$ (left panel) and $\pi^-$ (right panel) production in $\ell \, \pup \to \pi \, X$ processes, compared with the anti-tagged HERMES data~\cite{Airapetian:2013bim} and adopting the quark Sivers functions of Ref.~\cite{Anselmino:2016uie}. Curves have the following meaning: blue dashed lines for the LO and red solid lines for the LO+WW contributions.
}
  \label{fig:SSA-pi-xf02_016}
\end{figure}

We have then checked the impact of these new parameterizations on the description of HERMES data. The main results are the following: the Collins contribution is practically negligible for the fully-inclusive data set and tiny, but slightly improving the description, for the anti-tagged data category; while still confirming the good agreement with the fully-inclusive data, the use of the new Sivers parametrization reduces significantly the discrepancies between the theoretical predictions and the anti-tagged data for $\pi^+$ production (slightly overestimated adopting the SIDIS~1 and SIDIS~2 sets, see Fig.~\ref{fig:SSA-pi-xf02}, upper panels). In Fig.~\ref{fig:SSA-pi-xf02_016} we present the comparison of these new estimates with the anti-tagged data, noticing that, even at LO, one gets a clear improvement in the description of $\pi^+$ data. No significant differences appear in the unpolarized cross sections, where once again the WW piece is comparable with, or even dominates, the LO contribution.

Analogous features show up also in the predictions for JLab and COMPASS kinematics: almost no differences appear in the unpolarized cross sections, while a reduction in size of the SSAs for $\pi^+$ production (roughly a factor of two for COMPASS and three for JLab w.r.t.~SIDIS~2 estimates) comes out, leading to values for JLab at large $x_F$ of around -10\%. For $A_N$ in jet and $\pi^0$ production at EIC we find similar behaviours as those obtained adopting the SIDIS~1 set (see Fig.~\ref{fig:SSA-eic-pi}), with a reduction of our estimates by a factor of 1.5 at fixed $x_F$ vs.~$P_T$  (that is $A_N({\rm jet})$ is around 2\% at small $P_T$ and 1\% at the largest $P_T$ values) and by a factor of about three at fixed $P_T$ vs.~$x_F$. We have to notice that even if for the $P_T$ dependence this reduction could make the measurement of this asymmetry less feasible, the corresponding uncertainty band still presents a flat behaviour, implying a non vanishing and persisting SSA at large $P_T$. Concerning the $x_F$ dependence at large $x_F$ values one has to take into account the poor knowledge on the Sivers function in the large-$x$ region, heavily affecting also the new extraction.

Some general comments on these results are mandatory: $i$) the new extraction of the Sivers function (the dominant piece) is reasonably under control for HERMES kinematics. On the other hand some of the assumptions behind it (like the very different Gaussian widths in the unpolarized TMDs), still under current investigation, could have a non negligible impact on the predictions at JLab and EIC (see the above comments on the reduction factors); $ii)$~At the present stage it is then worth keeping and checking also the results obtained adopting the former fits, SIDIS~1 and SIDIS~2, because they are representative of different behaviours in the large $x$ region, still undetermined, and of different assumptions in their extractions; $iii)$ Even if these new extractions seem able to describe HERMES data quite well in a LO approximation, one has to keep in mind that in such a kinematical region the events are strongly dominated by the quasireal photon exchange contribution (see Section~\ref{unp-hermes}). The fact that the WW piece together with these new extractions gives a very good description of all HERMES data is the most interesting aspect of these results.

\section{Conclusions}
\label{concl}

SSAs observed in single-inclusive processes, like those measured in $p^\uparrow p\to h\, X$, where only one large energy scale is detected, represent a challenging issue in pQCD. Indeed, despite the rich amount of experimental data and their peculiar features, persisting up to the highest available energies, a thorough phenomenological description is so far missing and many theoretical aspects are still controversial and under debate.

Two approaches are nowadays adopted to describe these SSAs: one based on higher-twist parton correlation functions within a proven collinear factorization framework, and one based on transverse momentum dependent distributions within a phenomenological TMD scheme. Even if not formally proven, this last one enjoys quite considerable phenomenological successes and it is then worth to exploiting it further.

Because of the more complicated nature of SSAs in $pp$ collisions, and the difficulty in understanding their source, a study of SSAs in the theoretical more simple inclusive lepton-nucleon scattering processes, within a TMD scheme, was proposed in Ref.~\cite{Anselmino:2009pn} and then analysed against the available data in Ref.~\cite{Anselmino:2014eza}. These processes, moreover, share strong analogies with the SIDIS process for which TMD factorization has been proven. For these reasons they represent an important testing ground for the understanding of the origin of SSAs.

To assess the validity of the TMD scheme, the single spin asymmetry $A_N$, for the $\ell \, \pup \to h \, X$ process, was calculated in a leading-order approximation, adopting the Sivers and the Collins functions as extracted from SIDIS and $e^+e^-$ data.
Doing so, a unified TMD factorized approach is adopted, valid for $\ell \, p \to \ell' \, h \, X$ and $\ell \, p \to h \, X$ processes, in which, consistently, we obtain information on the TMDs and make predictions for $A_N$.

In the present analysis we have extended this strategy, including the contribution of quasireal photon exchange, in the Weizs\"acker-Williams approximation, expected to be important when the final lepton is scattered at small angles. To this aim we have calculated, for the first time, all involved TMD contributions both to the unpolarized and transversely polarized cross sections. We have then discussed the role of the WW term in the estimates of the unpolarized cross sections in various experimental setups, showing that it can be extremely important. In particular, at not so large $P_T$ values (like those explored at HERMES) and at large energies (as those reachable at an EIC) it could be comparable or even dominant w.r.t.~the LO contribution.

More important, we have shown how the description of the available SSA data from the HERMES Collaboration is significantly improved when the WW piece is included. Within the present approach we have also analysed the anti-tagged data events, not considered in the previous LO study. Even in this case our theoretical estimates show a good agreement with the data. It is also worth mentioning that the very few discrepancies in the description of some data sets (namely $\pi^+$ anti-tagged data) seem to disappear when adopting a very recent extraction of the Sivers distributions. This is another successful aspect of the phenomenological consistency of the entire approach. These are, in fact, the main findings of this study.

The role of the gluon Sivers function, which enters through the WW contribution, has been also investigated. Adopting the present knowledge on this TMD, even if with some caution, we have checked that its effect is negligible in most kinematical regions (at least where its extraction is constrained) and does not spoil the agreement with data. Further study would be nevertheless helpful.

Some predictions for ongoing or future experiments have been presented, pointing out the importance of new measurements in testing the overall picture. Notice that in many kinematical configurations the complete (LO+WW) calculation of the SSAs shows a very similar behaviour, in size and shape, as for the LO contribution. On the other hand, as extensively discussed, the WW piece changes significantly the expected yields of inclusive particle production.

Among the interesting perspectives of this study, we emphasize that at EIC, within a TMD scheme, one would expect similar features as those observed in the SSAs for $p^\uparrow p\to \pi\, X$ processes: the rising of $A_N$ with $x_F$ at fixed $P_T$, its almost vanishing at negative $x_F$ values and, somehow surprisingly, a flat behaviour of $A_N$ as a function of $P_T$. The very interesting case of SSAs in inclusive jet production, for which the Collins effect plays no role, has also been discussed, showing similar features as those for the inclusive neutral pion production.

This analysis could be definitely considered a further step towards a deeper understanding of the origin of SSAs in inclusive processes, and more generally, towards a unified TMD picture of these observables. All these findings, although quite encouraging, require further dedicated studies, both on the experimental and the theoretical side.

\acknowledgments
We thank M.~Schlegel and W.~Vogelsang for useful and interesting discussions on the proper use of the Weizs\"acker-Williams approximation in this context and M.~Anselmino for his careful reading of the manuscript.

\appendix
\section{Helicity formalism}
\label{appendix}

We collect here some details useful for the computation and the understanding of the expressions given in Section~\ref{formWW} (see Refs.~\cite{Anselmino:2004ky,Anselmino:2005sh} for a complete treatment).

We start recalling the helicity density matrix of a quark $q$, which can be written in terms of the quark polarization vector components, $\bm{P}^q = (P^q_x, P^q_y,P^q_z)$, defined in the quark helicity frame, as
\be
\rho_{\lambda^{\,}_q, \lambda^{\prime}_q}^{q/p,S} =
{\left(
\begin{array}{cc}
\rho_{++}^{q} & \rho_{+-}^{q} \\
\rho_{-+}^{q} & \rho_{--}^{q}
\end{array}
\right)}_{\!\! p,S} = \>
\frac{1}{2}\,{\left(
\begin{array}{cc}
1+P^q_z & P^q_x - i P^q_y \\
 P^q_x + i P^q_y & 1-P^q_z
\end{array}
\right)}_{\!\!p,S}\,.
\label{rho-q}
\ee
For a gluon (or any spin-1 massless particle) one can define the helicity density matrix as
\be
\rho_{\lambda_g^{\,}, \lambda^{\prime}_g}^{g/p,S}=
\frac{1}{2}\,{\left(
\begin{array}{cc}
1+P_{z}^{g} &
\mathcal{T}_1^g -i\mathcal{T}_2^g \\
\mathcal{T}_1^g +i \mathcal{T}_2^g & 1-P_{z}^{g}
\end{array}
\right)}_{\!\!p,S} \!\!\!\!\!\!\!
=
\frac{1}{2}\,{\left(
\begin{array}{cc}
1+ P^g_{circ}&
- P^g_{lin} \, e^{-2i\phi}\\
- P^g_{lin} \, e^{2i\phi} & 1-P^g_{circ}
\end{array}
\right)}_{\!\!p,S} \label{rho-gl} \!.
\ee
Eq.~(\ref{rho-gl}) refers, in general, to a mixture of circularly and linearly
polarized states. $P_{circ}^g$ corresponds to $P_{z}^g$, the gluon longitudinal
polarization. The off-diagonal elements of Eq.~(\ref{rho-gl}) are related to
the linear polarization of the gluons in the $(xy)$ plane at an angle $\phi$
to the $x$-axis.

Concerning the fragmentation sector we have, for a spinless (or unpolarized) hadron
\be
\sum_{\lambda_h }D_{\lambda_c,\lambda_c'}^{\lambda_h,\lambda_h}(z,\bm{p}_\perp) = D_{\lambda_c,\lambda_c'}^{h/c}(z,\bm{p}_\perp) = D_{\lambda_c,\lambda_c'}^{h/c}(z,p_\perp) e^{i(\lambda_c-\lambda_c')\phi_c^h}\,.
\ee
In particular for the quark fragmentation we have
\bea
&& \hat D_{++}(z,\bm{p}_{\perp}) = \hat D_{--}(z,\bm{p}_{\perp}) =
D_{h/q}(z,p_{\perp}) \\
&& \hat D_{+-}(z,\bm{p}_{\perp}) = D_{+-}(z,p_{\perp}) \, e^{i\phi_q^h} =
\frac{i}{2}\Delta^N\! D_{h/\qup}(z,p_{\perp})\, e^{i\phi_q^h} \,,
\eea
and for the gluon case
\bea
&& \hat D_{++}(z,\bm{p}_{\perp}) = \hat D_{--}(z,\bm{p}_{\perp}) =
D_{h/g}(z,p_{\perp}) \\
&& \hat D_{+-}(z,\bm{p}_{\perp}) = D_{+-}(z,p_{\perp}) \, e^{2i\phi_g^h} =
\frac{1}{2}\Delta^N\! D_{h/\mathcal{T}_1^g}(z,p_{\perp})\, e^{2i\phi_g^h} \,.
\eea

The remaining pieces to be considered are the helicity scattering amplitudes. The
transformations (a boost and two rotations) connecting the $p$-$\ell$ \emph{c.m.}~frame to the canonical $a$-$\gamma$ \emph{c.m.}~frame introduce some non trivial phases in the helicity amplitudes $\hat M_{\lambda^{\,}_c, \lambda^{\,}_d; \lambda^{\,}_a, \lambda^{\,}_\gamma}$, which
are a direct consequence of the non planar kinematics.

For massless partons there are only three independent elementary canonical amplitudes $\hat M^0$, corresponding to the $a\gamma\to cd$ processes we are interested in. This allows us to adopt the following notation
\be
\hat M_{++;++}  \equiv  \hat M_1^0 \,e^{i\varphi_1}\>\>\>\>
\hat M_{-+;-+}  \equiv  \hat M_2^0 \,e^{i\varphi_2}\>\>\>\> \label{phases}
\hat M_{-+;+-}  \equiv  \hat M_3^0 \,e^{i\varphi_3}\,,
\ee
where  $\hat M_1^0$, $\hat M_2^0$ and $\hat M_3^0$ are defined as
\bea
\hat M^0_{+,+;+,+} &=& \hat M^0_{-,-;-,-} \equiv \hat M_1^0 \nonumber\\
\hat M^0_{-,+;-,+} &=& \hat M^0_{+,-;+,-}\equiv \hat M_2^0 \nonumber\\
\hat M^0_{-,+;+,-} &=& \hat M^0_{+,-;-,+}\equiv \hat M_3^0 \>, \label{Mqq}
\eea
and the phases $\varphi_1$, $\varphi_2$ and $\varphi_3$ can be found in Refs.~\cite{Anselmino:2004ky,Anselmino:2005sh}.
Notice that the $+$ and $-$ subscripts refer to $(+1/2)$ and $(-1/2)$ helicities for quarks, and to $(+1)$ and $(-1)$ helicities for gluons/photons.

\end{document}